\documentclass[aps,twocolumn,superscriptaddress,floatfix,longbibliography,notitlepage]{revtex4-1}
\usepackage{graphicx}
\usepackage[export]{adjustbox}
\usepackage{amsmath,amssymb,wasysym}
\usepackage{braket}
\usepackage{mathtools}
\usepackage{mathrsfs}
\usepackage{verbatim}
\usepackage{makecell}
\usepackage{cases}
\usepackage{bbm}
\usepackage{enumitem}
\usepackage[dvipsnames]{xcolor}
\usepackage{hyperref}
\usepackage{empheq}

\newcommand*\widefbox[1]{\fbox{\hspace{2em}#1\hspace{2em}}}

\newcommand{\im}{\mathbf{i}}

\newcommand{\BigO}{\mathcal{O}}

\definecolor{Blue}{rgb}{0.00, 0.00, 1.00}
\definecolor{Red}{rgb}{1.00, 0.00, 0.00}

\newcommand{\bea}{\begin{eqnarray}}
\newcommand{\eea}{\end{eqnarray}}
\newcommand{\be}{\begin{equation}}
\newcommand{\ee}{\end{equation}}
\newcommand{\bee}{\begin{equation*}}
\newcommand{\eee}{\end{equation*}}

\begin{document}

\title{Clusters in an epidemic model with long-range dispersal}
\author{Xiangyu Cao}
\author{Pierre Le Doussal}
	 \affiliation{Laboratoire de Physique de l'\'Ecole normale sup\'erieure, ENS, Universit\'e PSL, CNRS, Sorbonne Universit\'e, Universit\'e Paris Cit\'e, F-75005 Paris, France}

\author{Alberto Rosso}
 \affiliation{Universit\'e Paris-Saclay, CNRS, LPTMS, 91405, Orsay, France}
 
 \begin{abstract}
   In presence of long range dispersal, epidemics spread in spatially disconnected regions known as clusters. Here, we characterize exactly their statistical properties in a solvable model, in both the supercritical (outbreak) and critical regimes. We identify two diverging length scales, corresponding to the bulk and the outskirt of the epidemic. We reveal a nontrivial critical exponent that governs the cluster number, the distribution of their sizes and of the distances between them. We also discuss applications to depinning avalanches with long range elasticity. 
 \end{abstract}
 
 \maketitle

Catastrophic events such as {avalanches, material failure, and initial-stage epidemic outbreaks}, often occur as a chain reaction. Their simplest model was that of Bienaym\'e and  Galton-Watson (BGW)\cite{Bienayme,WG}, {originally conceived for genealogy}. In a continuous time version one starts with a single infected individual.
During a short time lapse $\mathrm{d} t$ each infected individual recovers with probability $\gamma \mathrm{d} t$, and causes a new infection with probability $\beta \mathrm{d} t$. On average, each infection generates $R_0 = \beta / \gamma$ new ones: $R_0$ determines the fate of the epidemic. When $R_0 < 1$, it goes to extinction rapidly. When $R_0 > 1$, the size of  the population that has been  infected up to time $t$ grows exponentially, $ S \sim e^{(\beta-\gamma)t}$, as in the initial outbreak stage of an epidemic. At the critical point, $R_0 = 1$, the probability that the epidemic has survived up to time $t$ decreases as $\sim 1/t$, and in that case it will have infected $\sim t^2$ individuals. As a result, $S$ has strong  fluctuations and has a power law distribution $P(S) \sim S^{-3/2}$ with a cutoff at $S_{\rm max} \sim t^2$. The critical case mimics the scale free behaviour displayed by avalanches in disordered materials, i.e. the propagation of an instability which triggers further instabilities via elastic interaction~\cite{abbm}. 

The BGW model ignores the spatial spreading of the epidemic. Branching diffusion models consider that infected individuals also perform some random walk in a $d$ dimensional space, independently of recovery and infection.
Often, one specifies the random walk to be a short-range Brownian motion. Then the region affected by the epidemic is a connected set, whose geometric properties have been characterized~\cite{bramson,slade2002scaling,Brunet_2009,arguin,dumonteil13extent,ramola,ramola2}. 
For instance, at criticality, the radius $\xi$ of this set grows as $\xi \sim S^{1/4}$.
\begin{figure}
    \centering
    \includegraphics[width=.85\columnwidth]{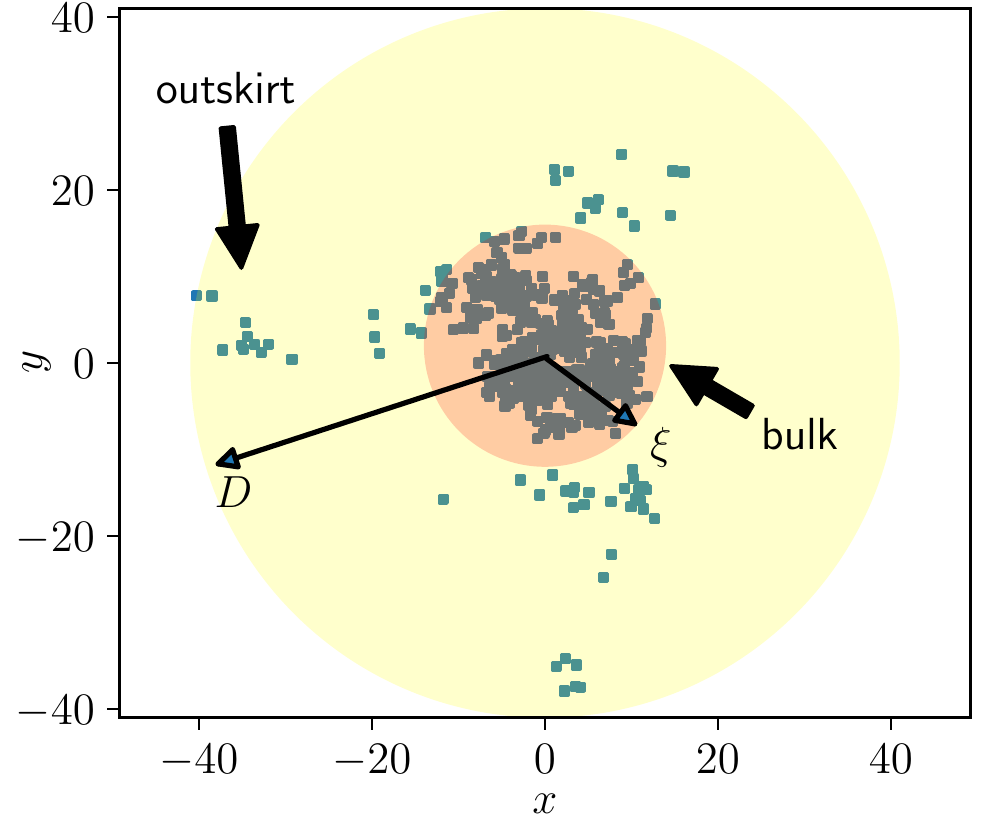}
    \caption{Spatial distribution of an critical epidemic started at the origin, totaling $1000$ infections. Due to the long-range dispersal [\eqref{eq:pr}, $\alpha=1.5$] of infected individuals, the points visited form disconnected clusters. The bulk of radius $\xi$, concentrating a majority of infections, is surrounded by a sparse outskirt, containing \textit{all} infections.   }
    \label{fig:picture}
\end{figure}

However, Brownian diffusion models cannot capture the \textit{long-range} dispersal that ubiquitously occurs in nature, due to e.g. wind, ocean currents, and air traffic~\cite{01ants,02aerial,nathanplant,brockmann,gonza08nature,perlekar10turbulent,barrat}, spreading an epidemic far from its origin. A similar situation is observed in disordered materials where long-range interactions can trigger disconnected avalanches, e.g., in the propagation of crack fronts~\cite{rice85,gao89crack,tanguy98crack,bonamy08crack}, wetting lines~\cite{joannydegennes,moulinet04,LeDoussal2009} or plasticity~\cite{roux,linjie}.  In this work we model the long-range dispersal of the infected individuals as follows: during $\mathrm{d} t$, an individual jumps from $x$ to $x'$ with probability $p_\alpha(x-x') \mathrm{d}^d x' \mathrm{d} t$, where $p_\alpha(x)$ decays as a power law at large distances:%~\footnote{}:
\begin{equation}\label{eq:pr}
p_\alpha(x) = \frac{\theta(|x| - \epsilon)}{|x|^{\alpha + d}}  \,,\, \alpha > 0 \,.
\end{equation}   
Here {$|x|$ is the Euclidean norm}, $\theta$ is the Heaviside step function and $\epsilon \ll 1$ is a short-distance cutoff. Similar long-range models have been studied {on a lattice}, where the outbreak always displays a sub-exponential growth~\cite{doi:10.1073/pnas.1404663111,chtterjee,caobouchaud,hinrichsen,jansen08DP,Grassberger13-1d,grassberger13-2d}. Here, we assume an infinite pool of susceptible individuals everywhere, which ensures an exponential outbreak when $R_0>1$. 

A typical epidemic obtained from a numerical simulation of our model is shown in Fig.~\ref{fig:picture}. One may distinguish two regions characterized by distinct length scales. The {\it bulk}, of radius $\xi$, contains most of the infections. Farther away, a sparser {\it outskirt} of radius $D$ contains all the remaining infections. The existence of the outskirt is a consequence of the long-range jumps. One aim of this work is to obtain how $\xi$ and $D$ scale with the infected population $S$. 
Another fundamental consequence of long-range dispersal, is the presence of \textit{clusters}, i.e. spatially disconnected regions affected by the epidemic. As is apparent from  Fig.~\ref{fig:picture}, the clusters vary in sizes and their spatial distribution is not uniform. The second goal of this Letter is to introduce a method to properly define the clusters. We then characterize their random geometry: how the number of clusters grows with $S$, how their sizes are distributed, what are the distances separating them, etc. Our exact results are obtained by the analysis of a non-linear "instanton" equation. {We stress that our methods are applicable to real-world data. As a proof of principle, we tested our theory against the Covid-19 outbreak data in the US. Remarkably, a prediction of our model, \eqref{eq:NgSC} below, describes well the spatial distribution of the clusters during the first week of March 2020~\cite{SM}.}
% --- applicable to real-world data ---

 %
The epidemic model introduced above provides a discrete realization, equivalent, near criticality~\cite{pierrenew}, to the mean-field theory ~\cite{pldw12,pldw13} describing the spatial structure of the avalanches of slowly driven elastic interfaces in a disordered medium. 
In crack experiments, clusters have been directly observed~\cite{toussaint06},
and their number and size distribution have been characterized~\cite{zapperi10,lepriol21}. These works proposed that these properties are fully encoded in the global properties of the crack front, e.g., in its roughness exponent~\cite{kardar94,twoloop,rosso02rough}. Here, we make a first step at examining this issue analytically; our results indicate that the cluster statistics probably involve a new independent exponent. In what follows, we report our main results, and sketch the main points of their derivation, see~\cite{SM} for details.

\noindent\textit{Bulk and outskirt.}  We first determine the length scales of the bulk, $\xi$, and outskirt, $D$, by simple arguments. We consider our model with a single infected individual at the origin initially ($t=0$).  At criticality ($R_0 = 1$), the bulk length $\xi$ can be estimated as the \textit{typical} displacement of a random walk with jump distribution \eqref{eq:pr}. When $\alpha < 2$, we have a L\'evy flight, and thus
 \begin{equation}\label{eq:bulksize}
    \xi \sim t^{\frac1\alpha} \sim S^{\frac1{2\alpha}} \,,\, \alpha < 2 \,,
\end{equation}
where the last estimate comes from the scaling $S \sim t^2$. When $\alpha > 2$, we recover the short-range behavior $\xi \sim \sqrt{t} \sim S^{\frac14}$~\cite{pldw13,Thiery_2015}. 
On the other hand, the outskirt's diameter $D$ is estimated as the \textit{farthest} jump among $\sim S$ independent attempts:
\begin{equation}
    D \sim S^{\frac1{\alpha}} \,.  \label{eq:D}
\end{equation}
Hence, the outskirt is much larger than the bulk if $\alpha < 4$: only for $\alpha > 4$ do we completely recover a short-range behavior, with $D \sim \xi \sim S^{1/4}$. This is already a surprise, as naively one would expect a short-range takeover at $\alpha = 2$. 

In the supercritical regime, the argument for the outskirt diameter $D$ and the result~\eqref{eq:D} still hold. The scaling of the bulk size $\xi$ is different. Indeed, the infected population grows exponentially, $S\sim e^{(\beta-\gamma)t}$. As a consequence, the density of infected individuals is exponentially large at the epicenter, $x=0$, and decays as $\sim S |x|^{-\alpha-d}$~\eqref{eq:pr}. The bulk extent is then determined by the distance, $|x|=\xi$, at which the density reaches unity:
\begin{equation}\label{eq:bulk-SC}
    \xi \sim S^{\frac{1}{\alpha + d}} \sim e^{\frac{\beta-\gamma}{\alpha+d} t}
\end{equation}
Note that, when $R_0>1$, the separation of scales $D\gg \xi$ remains for {any} $\alpha$ and the short-range behavior with a linear growth $\xi \propto t$ is never recovered. This is in contrast with lattice models~\cite{chtterjee,doi:10.1073/pnas.1404663111}, where a reduction to short-range does happen at $\alpha = d+1$. % 
 \begin{figure}
    \centering
    \includegraphics[width=.85\columnwidth]{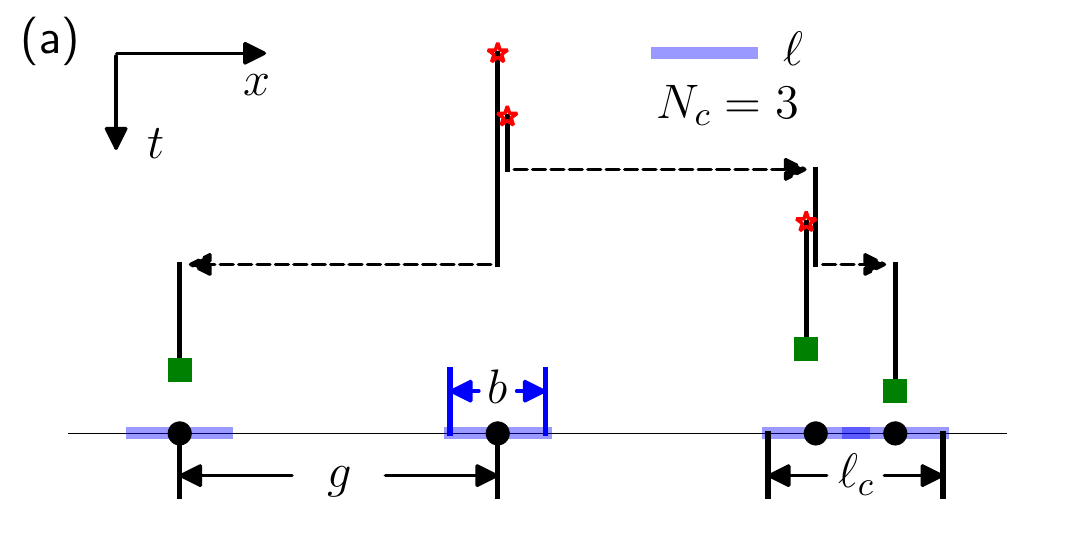}
      \includegraphics[width=.85\columnwidth]{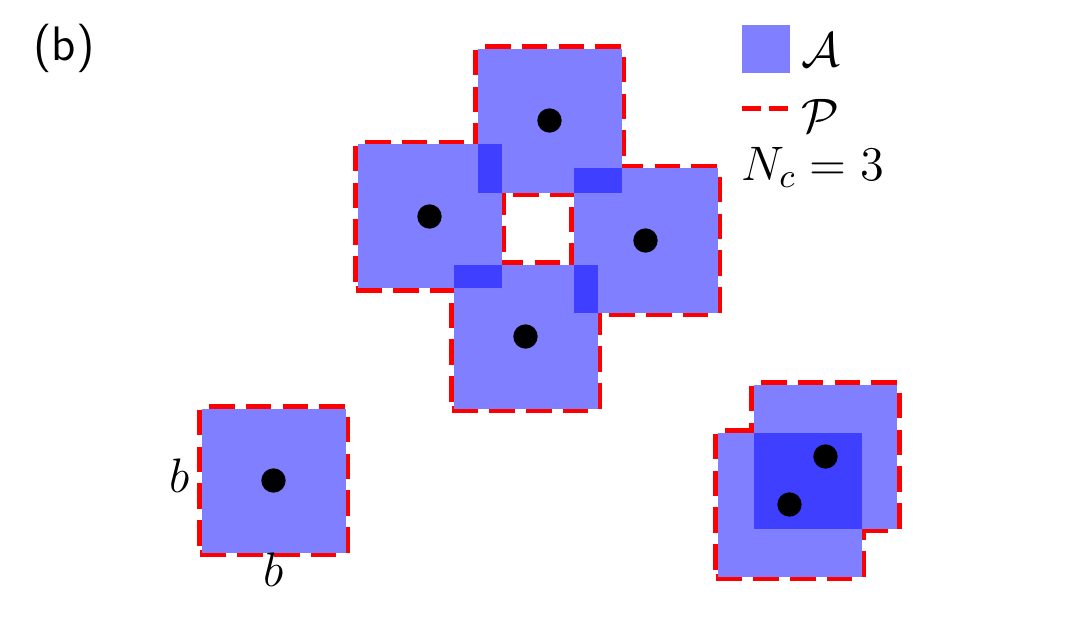}
    \caption{(a) Illustration of an epidemic in 1D. An infection (recovery, jump) is indicated by a red star (green square, dashed line, resp.). The points visited are coarse-grained by an interval of length $b$. They form $N_c=3$ clusters, with total extension $\ell=\sum \ell_c$. The gaps $g$ are defined independently of $b$. (b) In 2D, a point is coarse-grained by a square of side $b$, to define the cluster number $N_c (= 3)$, the area $\mathcal{A}$, and the perimeter $\mathcal{P}$. Note that in 2D, a cluster can be non-convex and have holes. }
    \label{fig:defs}
\end{figure}

\noindent\textit{Defining clusters}. In our model, the ensemble of positions ever occupied by an infected individual up to time $t$ is a finite set, as only a finite number of jumps have occurred. How do we define its clusters? For simplicity we focus on one and two dimensions. We introduce a coarse-graining scale $b \gg \epsilon$, and thicken each point by a patch of size $b$ --- an interval of length $b$ in 1D, and a square of size $b$ in 2D --- centered at that point, see Fig.~\ref{fig:defs}. The patches attached to different points can then overlap and form clusters. {To characterize {their spatial distribution}, we introduce the following observables: (i) The number of cluster $N_c$; (ii) the length/area of individual clusters, $\ell_c$ in 1D and $\mathcal{A}_c$ in 2D. The sum of all $\ell_c$ ($\mathcal{A}_c$) is the epidemic's \textit{extension}, $\ell$ (area, $\mathcal{A}$, respectively). (iii) We also characterize the distances between clusters. In 1D, a natural choice is the distribution of gaps (Fig.~\ref{fig:defs}). It is not hard to see that, the number of gaps larger than $g$ is related to the cluster number with $b = g$: 
\begin{equation}
    N_c(b=g) = \text{(number of gaps $> g$)} + 1 \,.
\end{equation}
In 2D, the notion of gaps is not obvious, and we take $N_c(b=g)$ as a probe of the distances between clusters. We obtained the $b$ dependence of all the quantities; for conciseness, we report results with $b=1$ unless otherwise stated.}

\noindent\textit{Clusters at criticality.} { When $R_0 = 1$, statistical fluctuations are strong, and there are various ways of averaging. Here, we focus on averages conditioned on a large infected population $S$ {  (assuming non-extinction)}, denoted as $\left< \mathcal{O} \right>_S$ for an observable $\mathcal{O}$. From the $S$-conditioned averages, we can obtain the asymptotics of the average over all realizations up to time $t$, using 
\begin{equation}
    \left< \mathcal{O}(t) \right> \sim \int^{S_{\max} } P(S)
\left< \mathcal{O} \right>_{S} \mathrm{d} S,\quad S_{\max} \sim  t^2 \,.
\end{equation}  
Thus, if $\left< \mathcal{O} \right>_{S} \sim S^{a}$, $\left< \mathcal{O}(t) \right> \sim t^{\max(2a-1, 0)}$ (See Table I of \cite{SM} for results). }
%}

We have seen that when $\alpha < 4$, the outskirt is much larger than the bulk, and we expect many clusters. Interestingly, the interval $\alpha \in (0,4)$ is divided into several regimes, with qualitatively different behaviors of $\left< \ell \right>_S, \left< \mathcal{A} \right>_S$ and $\left< N_c\right>_S$. Let us start with the most nontrivial one, $\alpha \in (d/2,d)$. 
{ There, we find that the average extension and area are related to the bulk extent in a rather expected way:
\begin{equation}\label{eq:ellandA}
    \left< \ell \right>_S \sim  \xi \,,\, 
    \left< \mathcal{A} \right>_S \sim \xi^2   \,.
\end{equation}
It is worth noting that the above quantities are independent of $b$ for a large range of $b$, see \eqref{eq:ellb_main} below.} Now, the average number of clusters scales with $\xi$ via a new and nontrivial exponent
\begin{equation}
 \left< N_c \right>_S \sim \xi^{\chi}  \,,\, \alpha < \chi <d  \,. \label{eq:NcS}
\end{equation}
The exponent $\chi$ is a function of $\alpha$ and $d$, and determined by a transcendental equation given in \cite{SM} together with a plot. It satisfies $\alpha < \chi < d$, which means that the number of clusters grows with $S$ but remains much lower than the area or extension. Thus, the cluster areas $\mathcal{A}_c$ and extensions $\ell_c$ must have broad distributions (with divergent mean as $S\to\infty$). Computing them is beyond the reach of the present techniques. However, assuming that they follow a single power law in the interval $[1, \xi^d]$, we can surmise their exponent~\cite{lepriol21}: 
\begin{equation}
  P(\ell_c) \sim \ell_c^{-\chi-1} \,,\, 
   P(\mathcal{A}_c) \sim \mathcal{A}_c^{-\chi/2-1} \,.\,
   \label{eq:conjecture}
\end{equation}
 
\begin{figure}
    \centering
    \includegraphics[width=.95\columnwidth]{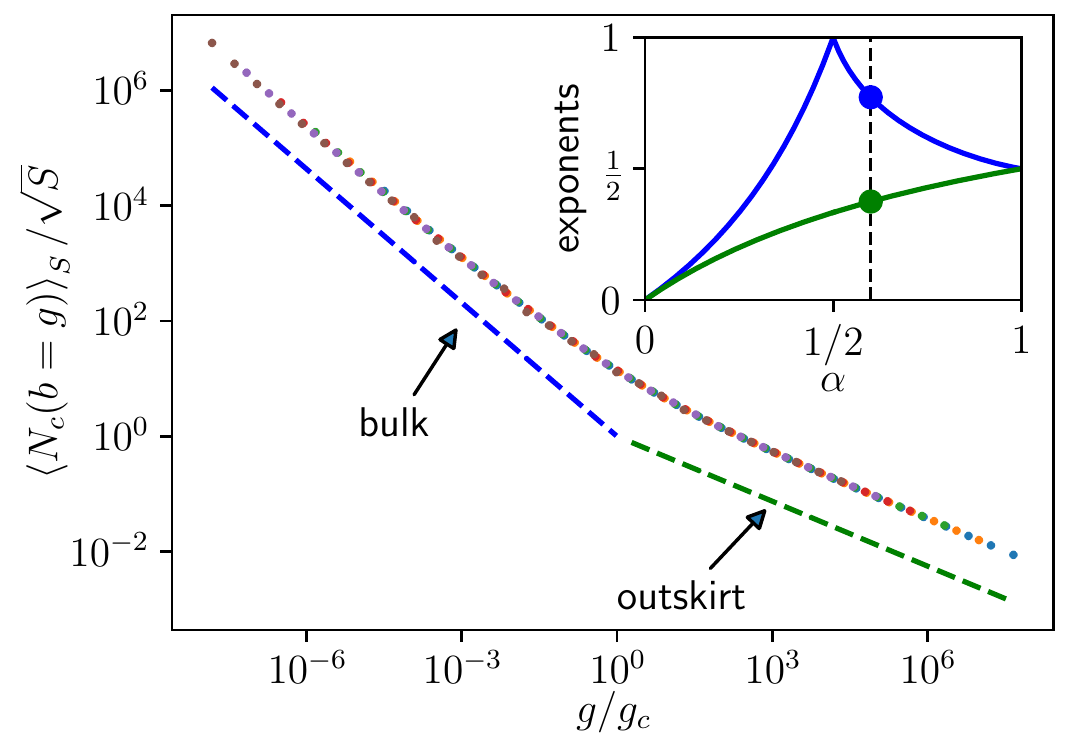}
    \caption{Gap distribution in 1D with $\alpha = 0.6$, obtained by numerical solution of~\eqref{eq:travellingwave}\cite{SM}. Data points with various sizes $S = 10^{20}, \dots, 10^{28}$ are collapsed using \eqref{eq:gapS}. The dashed lines indicate the predicted exponents in two regimes. Inset: The dependence of the two exponents on $\alpha$. }
    \label{fig:gap}
\end{figure}

{Concerning the {gaps between clusters}, we found that $ \left< N_c(b = g) \right>_S$ has two regimes with distinct power laws:  
\begin{equation}\label{eq:gapS}
 \frac{ \left< N_c(b = g) \right>_S}{\sqrt{S}} \sim 
    \begin{dcases} 
      (g/g_c)^{-\frac{d(\chi-\alpha)}{d-\alpha} } &  1 \ll g \ll g_c  \\
        (g/g_c)^{-\frac{\alpha d}{d+\alpha}}  & g_c \ll g \ll  D
    \end{dcases} \,,
\end{equation}
where $g_c = \xi^{1- \alpha / d }$ is the crossover gap length. The two gap regimes $g \ll g_c$ and $g \gg g_c$ correspond to gaps in the bulk and in the outskirt, respectively. To better understand this result, let us consider one dimension~\footnote{A similar but less precise description applies to 2D if we replace a gap of length $g$ by an ``empty space'' of area $g^2$.}. Observe that, the total length of the gaps no greater than $g_c$ is exactly the bulk size: \begin{equation}
\sum (\text{gaps  $\le g_c$}) \sim g_c \left< N_c(b = g_c) \right>_S \sim \xi \,. \end{equation}
Now, if we consider all the bulk gaps up to a size $g \ll g_c$, their number is almost $N_c(b=1)$, but their total size is a negligible fraction of the bulk. On the other hand, the outskirt gaps of are a minority in number, but their total size is much greater than the bulk size.} Of course, there is no sharp transition between bulk and outskirt, but rather a smooth crossover. Indeed, in Fig.~\ref{fig:gap}, we show that to demarcate the two power laws requires several orders of $g/g_c$. {  Otherwise, one may observe a ``compromise'' of the theoretical predictions.} %Monte Carlo studies of this long-range model must be executed with great care. 

{ So far we focused on the regime $\alpha \in (d/2,d)$. The other ones are simpler. In a nutshell, for strong long-range dispersal ($\alpha < d / 2$), the clusters become atomic and have a finite size in average. Therefore we have $\left<N_c\right>_S \sim S$, and $\left<\ell\right>_S, \left< \mathcal{A} \right>_S \sim S$ as well. For weak long-range dispersal ($\alpha > d$), the bulk becomes more compact, and gaps of size $\gtrsim 1$  exist only in the outskirt. See~\cite{SM} for a detailed discussion. }

\noindent\textit{Clusters of an outbreak.} In the super-critical ($R_0 > 1$) regime, the statistical fluctuations are weak. We can thus consider the averages up to $t$, which are dominated by realizations with an infected population $S \sim e^{(\beta-\gamma)t}$. Recall that the bulk and outskirt diameter grow exponentially as \eqref{eq:D} and \eqref{eq:bulk-SC}, for any $\alpha > 0$. Now, the cluster structure of an outbreak is also simpler, and we found the same qualitative picture for any $\alpha$. The bulk is compact and has no large gaps. Its extension/are is $\left< \ell \right> \sim \xi, \left<\mathcal{A} \right> \sim \xi^2$, where $\xi \sim S^{1/(d+\alpha)} \sim e^{(\beta-\gamma)t/(d+\alpha)}$~\eqref{eq:bulk-SC}. The outskirt is sparse, and has an exponential number of clusters, $\left< N_c \right> \sim \xi^d$. Notably, their spatial structure is time-independent: the gap distribution is stationary up to a normalization and a cutoff,
\begin{equation}
  \left< N_c(b = g) \right> \sim \xi^{d} g^{-\frac{d \alpha}{\alpha+d}} \,,\, g \ll D \,. \label{eq:NgSC}
\end{equation}
{In~\cite{SM}, we tested this prediction against Covid-19 data, finding an encouraging agreement. }

\noindent\textit{Method.} We highlight some key points of our analytical approach. The main object is a function $F(x, t\vert b)$, which is the probability that $x$ belongs to the patch of some point visited before $t$. A standard backward recursion argument shows that $F$ satisfies a {semi-linear} ``instanton'' equation~\cite{fisher,kpp,pldw13,dawson,wanatabe}:
\begin{equation} \label{eq:travellingwave}
    \partial_t F = \mathcal{D}^\alpha F + (\beta-\gamma)F - \beta F^2 \,,\, F\vert_{t=0} = 0 \,,
\end{equation}
for any $x$ outside the patch of the origin; inside that, $F = 1$. Here $(\mathcal{D}^\alpha f)(x) := \int p_\alpha(x-y) (f(y) - f(x)) \mathrm{d}^d y$ is the ``fractional diffusion'' term. From the solution $F$, we can obtain the area (extension) by integrating it over the plane (line). The cluster number is obtained by differentiating with respect to $b$. In 1D, we have 
\begin{equation}
    N_c(b) = \partial_b \ell(b) \,. \label{eq:Ncellb}
\end{equation}
A similar trick exists in 2D~\cite{SM}. 

\begin{figure}
    \centering
    \includegraphics[width=.9\columnwidth]{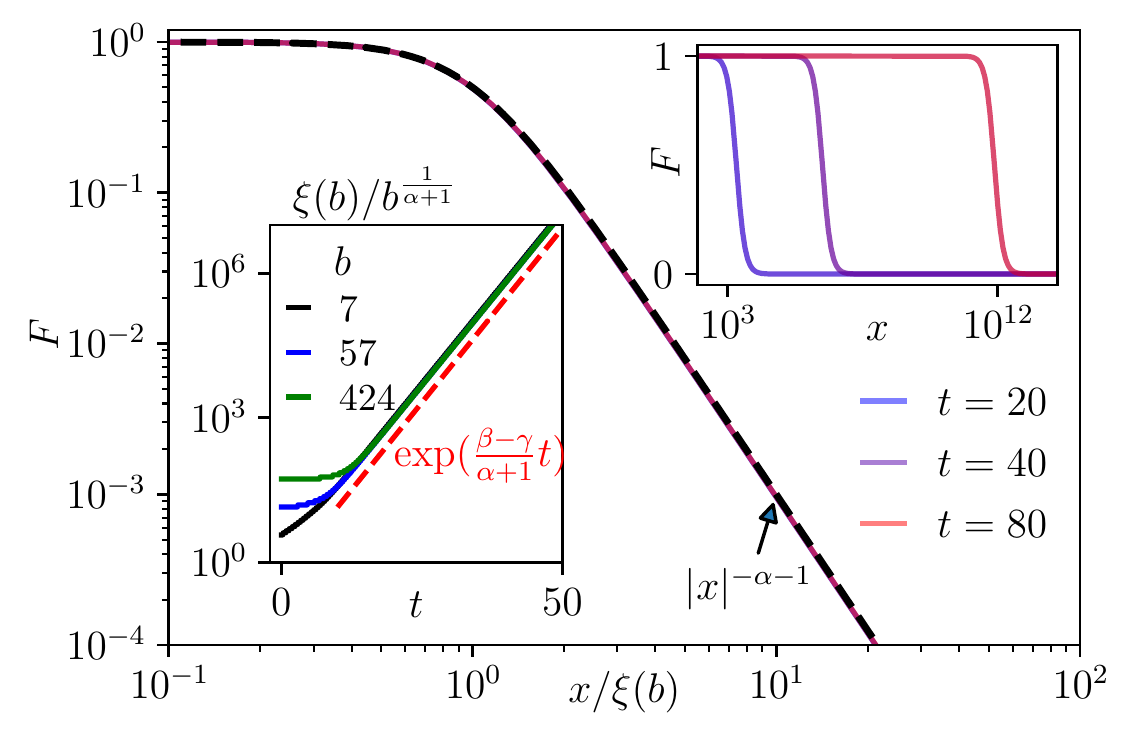}
    \caption{Traveling wave solution to \eqref{eq:travellingwave} in the supercritical regime, ($\alpha = 2, \beta =1, \gamma=0, b = 7$). The solution at $t=20,40,80$ (upper inset) collapsed onto the front profile $F(x,t) = f(x/\xi(b))$, $f(y) = 1/(1+y^{1+\alpha})$ (black dashed curve). The front position $\xi(b)$ is defined by $F(\xi(b)) = 1/2$. Its time dependence is plotted in the lower inset for 3 values of $b$. The collapse confirms the $b$ dependence of the front position~\cite{SM}.}
    \label{fig:travel}
\end{figure}
Therefore, the problem boils down to the asymptotic analysis of \eqref{eq:travellingwave}. In the super-critical regime, the exponential spreading of its traveling wave solution follows from existing rigorous results~\cite{Cabre13KPP}; for a self-contained derivation and our results on clusters, see~\cite{SM}. In Fig.~\ref{fig:travel}, we plot the front profile. Note that it decays as a power law, and does not have a characteristic width. In contrast, in traveling wave equations with short-range diffusion, the wavefront position has linear growth in time and its width is of order unity. 

The results at criticality follows from the stationary solution of ~\eqref{eq:travellingwave}. The solution in the regime $\alpha \in (d/2, d)$ involves a noteworthy feature. To discuss that without going into technical details, consider the following puzzle, say in 1D. Recall that the cluster number and the extension are related by a $b$-derivative~\eqref{eq:Ncellb}. Then, how can they scale differently: $\ell \sim \xi, N_c \sim \xi^\chi\ll \xi$? The crux is that, the leading asymptotics of $\ell$ is $b$-independent, while $N_c$ derives from a subleading term:
\begin{equation}\label{eq:ellb_main}
    \left< \ell(b) \right>_S =  c_0 \xi + c_1(b) \xi^{\chi}  \,,\, (b \ll g_c )
\end{equation}
where $c_0$ is $b$-independent. To extract the cluster statistics from the solution of~\eqref{eq:travellingwave}, it is necessary to identify its \textit{subleading} asymptotics, in addition to the previously known leading one~\cite{lepriol-thesis}. This mathematical detail has a physical interpretation: cluster statistics are associated with irrelevant perturbations in the sense of the renormalization group. During the coarse-graining process, the clusters merge and information about them is gradually erased.

\textit{Conclusion}. We have characterized the clusters of an epidemic model with long-range dispersal, which is equivalent near criticality to the mean-field theory of depinning avalanches with long-range elasticity. We found that two diverging length scales --- the bulk and the outskirt --- emerge in both super-critical and critical regimes. In the latter, the bulk can have a rich structure with broadly distributed cluster sizes as well as gap sizes. Our analytical approach based on the instanton equation can be extended to study the effect of {  inhomogeneous networks~\cite{brockmann}, realistic mixing patterns~\cite{brockmann,mistry21}, super-spreading events~\cite{lloyd05}}, or the regions where the epidemic is still active at time $t$~\cite{meyer96,Houchmandzadeh}. It will be also interesting to see how much the qualitative features revealed here appear in other epidemic models, e.g., contact point processes~\cite{marro_dickman_1999,krapivsky2010kinetic}.
Finally, concerning depinning avalanches, our {model provides a mean-field description which
should be quantitatively correct for realistic long-range systems when $d \geq 2\alpha$. To describe these systems for $d<2 \alpha$, loop corrections to mean field theory should be taken into account.} In particular, our results imply a cluster number distribution $P(N_c) \sim N_c^{- \mu}$ where $\mu={\alpha}/{\chi}+1$ for $\alpha \in (d/2,d)$. At the critical dimension $d=2 \alpha$, we recover the BGW value $\mu = 3/2$, but {in our model} $\mu > 3/2$ is a new exponent when $d< 2 \alpha$. Meanwhile, numerical studies~\cite{lepriol21} of realistic models suggest that $\mu\approx 3/2$ for  all $d<2\alpha$. {It will be interesting to see how to retain the ``dangerously irrelevant'' cluster statistics in the field theory and whether the loop corrections can account for this numerical observation}. 

\begin{acknowledgments}
We thank Jean-Philippe Bouchaud for pointing out the literature on human mobility. We thank William Terrot for preliminary work on the project, and Gr\'egory Schehr, Vincenzo Schimmenti for valuable comments on the manuscript. PLD acknowledges support from ANR under the grant ANR-17-CE30-0027-01 RaMaTraF. XC and PLD thank LPTMS for hospitality. 
\end{acknowledgments}

\pagebreak
\begin{widetext}

\section*{Supplemental Material}
\subsection{Derivation of the instanton equation}
We recall the standard backward recursion argument used to derive the instanton equation. Let the positions of the infected individuals at time $t$ be $x_1, \dots, x_{I(t)}$ where $I(t)$ is the number of infected. Note that at $t= 0$ we have one infected individual at $x_1 = 0$. Consider the probability that the $b$-neighborhood (called a patch in the main text) around $x$ has not been infected until $t$:
\begin{equation}\label{eq:E}
     E(x,t\vert b)=1 - F(x,t\vert b) = \mathrm{Prob}(  \Vert x_i(s) - x \Vert > b/2 , \forall i = 1, \dots, I(s), s < t ) 
\end{equation}
where $\Vert x \Vert = \Vert x \Vert_{\infty}$ is the infinite-norm of $x$, defined as $\Vert (x^1, \dots, x^d) \Vert = \max \{x^1, \dots, x^d\}$ ($x^a$ is the $a$-th component of a point $x$). In particular, $\{\Vert x \Vert < b/2\}$ is the box of linear size $b$ centered at the origin. 

Since the spatial diffusion is symmetric (the probability of going from $x\to y$ and $y\to x$ are the same), it is not hard to see that $E$ is equal to the probability that the $b$-neighborhood of the origin has not been visited, if the epidemic starts at $x$: 
\begin{equation}
       E(x,t\vert b) = \mathrm{Prob}(  \Vert x_i(s)\Vert > b/2  \vert x_0(0) = x ) \,. 
\end{equation}
Note that if $\Vert x \Vert \le b/2$, $E = 0$ by definition. For $\Vert x \Vert > b/2$, we can apply a backward recursion of $E$ by considering what can happen during $t \in (0,\mathrm{d}t)$. 
\begin{enumerate}
    \item Another individual is infected, with probability $\beta \mathrm{d} t$. In that case $E(x) \to E(x)^2$ (because from that moment, the two individuals act independently from now on with the same law).
    \item The patient 0 recovers with probability $\gamma \mathrm{d} t$. Then $E(x) \to 1$ (note that we assumed $\Vert x \Vert > b/2$). 
    \item The patient performs a jump to $y$ with probability $p_\alpha(y-x)\mathrm{d}^d y \mathrm{d} t$. In that case $E(x) \to E(y) $.  Note that $p_\alpha(x)$ is a probability rate, and thus not normalized. With probability $1-\mathrm{d} t \times \int p_\alpha(x) \mathrm{d}^d x $, the individual makes no jump.
\end{enumerate}
Gathering all the possibilities, we have
\begin{equation}
    E(x, t + \mathrm{d} t) - E(x, t) = \beta (E(x)^2 - E(x)) \mathrm{d} t + 
     \gamma (1 - E(x)) \mathrm{d} t + \int  p_\alpha(x-y) (E(y)-E(x)) \mathrm{d}^d y   \mathrm{d} t \,,\, \Vert x \Vert > b/2 \,.
\end{equation}
Now noting that $F = 1-E$ and the definition of the fractional diffusion operator
\begin{equation}
    (\mathcal{D}^\alpha f)(x) = \int p_\alpha(x-y) (f(y)-f(x)) \mathrm{d}^d y \,,\, p_\alpha(x) = |x|^{-d-\alpha} \theta(|x|-\epsilon) \,,
\end{equation}
we obtain the instanton equation, \eqref{eq:travellingwave-supp} below.

In the main text we mentioned that in 1D, the cluster number can be obtained by deriving the extension with with respect to $b$, namely $  N_c(b) = \partial_b \ell(b)$. In 2D, deriving the area once gives the \textit{perimeter} $\mathcal{P}$; deriving twice, we obtain the difference between cluster and hole numbers: 
\begin{equation}\mathcal{P}(b) = 2 \partial_b \mathcal{A}(b) \,,\, 
\partial_b \mathcal{P}(b) = 4 (N_c - N_h) \,.
\end{equation}
Yet, we are able to constrain the asymptotics of $N_c$ using the bounds 
\begin{equation}
\mathcal{P} / (4b) \ge N_c \ge N_c-N_h  \,.
\end{equation}
These geometric formulas are not hard to derive, upon observing Figure 2 of the main text. We also note that similar formulas (with different prefactors) hold if we replace squares by disks in 2D. So our asymptotic results are  independent of this choice.

\subsection{Super-critical regime: traveling wave solution}
We consider the instanton equation for $F(x,t\vert b)$, which is the probability that the $b$-neighborhood of $x$ has been visited by an infected individual by time $t$ (the epidemic starts with a single infected individual at $x=0,t=0$) 
\begin{align} \label{eq:travellingwave-supp}
    \partial_t F = \mathcal{D}^\alpha F + (\beta-\gamma)F - \beta F^2 \,,\, \Vert x \Vert > b/2  \\ 
    F(\Vert x \Vert < b/2) = 1 \,,\, F\vert_{t=0} = 0 \,,
\end{align}
in the super-critical regime ($\beta > \gamma$). Note that the initial condition $F\vert_{t=0} = 0$ is consequence of the strict inequality $s<t$ in the definition~\eqref{eq:E}: for $t = 0$, $F = 0$ because no infected individual exists for $t < 0$. 

Here, we provide a simple self-contained derivation of the traveling wavefront position (including $b$ dependence) and of front profile. The result applies to any $d$ and $\alpha$. Very initially, we can neglect the last two terms and solve the approximate equation $\partial_t F =  \mathcal{D}^\alpha  F\vert_{t=0}$. We get
\begin{align}
 & F \approx t \mathcal{D}^{\alpha}\left[\theta(b/2-\Vert x \Vert)\right] \,,\, \\ & \mathcal{D}^{\alpha}\left[\theta(b/2-\Vert x \Vert)\right]\approx \begin{cases}
      (\Vert x \Vert-b/2)^{-\alpha} &  0 < \Vert x \Vert-b/2 \ll b \\ 
     b^d | x |^{-\alpha-d} &   |x| \gg b \,. 
  \end{cases}\label{eq:Dplateau-sup}
\end{align}
The first regime corresponds to points outside the $b$-neighborhood but very close to it (such that the neighborhood appears semi-infinite). In that case there can be prefactors in the above formula depending on $x/b$, but they are unimportant for what follows. The second regime corresponds to points far away from the neighborhood; the formula we gave is asymptotically exact. 

At $t = \BigO(1)$ (uniformly for all $x$), the linear in $t$ growth is be overtaken by the exponential growth generated by the $(\beta-\gamma)F$ term. In fact that term dominates the RHS of \eqref{eq:travellingwave-supp}, so that 
\begin{equation}
    F \sim S \mathcal{D}^{\alpha}\left[\theta(b/2-\Vert x \Vert)\right] \,,\, \text{where } S = e^{(\beta-\gamma)t} \,, \label{eq:Fgrowth}
\end{equation}
\textit{until} $F\sim 1$ and the nonlinear term $-\beta F^2$ stops the growth. If $S \gg b^{\alpha}$, e.g. if the  scale $b$ is smaller than 
outskirt scale  $D =  S^{1/\alpha}$, we can obtain an equation for the wave front position $\xi(b)$
\begin{equation}
   S   \sim  b^{-d} \xi(b)^{\alpha+d}  \implies  \xi(b) = S^{\frac{1}{\alpha +d}} b^{\frac{d}{\alpha + d}} \,,\, \text{if }D \gg b \,.
\end{equation}
We remark that $\xi(b)$ is related to the bulk extent $\xi$ by the relation $\xi=  \xi(b=1)$, but they are different quantities. Recall that in the super-critical regime, we define the bulk extent as the distance from origin at which the density of the infected population becomes of order one. $\xi$ is thus independent of $b$, and can be done by a simple argument, as given in the main text, and does not require analyzing the  instanton equation. In contrast, $\xi(b)$ is the $b$-dependent wavefront position of the instanton equation.

Now, we can plug the traveling wave ansatz 
\begin{equation}
    F(x,t) = f(|x|/\xi(b)) 
\end{equation}  
into \eqref{eq:travellingwave-supp} to find the front profile. As a result, at large $\xi$, we find  
\begin{equation}
  -  \frac{(\beta-\gamma) y  f'(y)}{\alpha + d}  = (\beta-\gamma) f(y) - \beta f(y)^2  \,.
\end{equation}
Note that the $\mathcal{D}^\alpha F$ term gives a negligible contribution. We can explicitly solve for $f(y)$:
\begin{equation}
    f(y) = \frac{\beta-\gamma}{\beta + (y/y_0)^{d+\alpha}} \label{eq:frontprofile}
\end{equation}
where $y_0$ is an unknown constant. Note that $f(y \to 0) \to 1-\gamma/\beta = 1-R_0^{-1}$ and $f(y\to\infty) \sim 1 / y^{d+\alpha}$. These predictions are verified in Fig.~4 of the main text.

In summary we have shown that for any $b$ fixed, as $t\to\infty$, 
\begin{equation}
    F(x,t) \to f(|x|/\xi(b)) \,,\, \xi(b) = e^{\frac{\beta-\gamma}{\alpha +d} t} \, b^{\frac{d}{\alpha + d}} 
\end{equation}
with $f$ given by \eqref{eq:frontprofile}. The bulk extent is given by $\xi$ with $b = 1$. Integrating over $x$ in 1D and 2D, we have 
\begin{align}
    \left< \ell \right> = \int F \mathrm{d} x \sim \xi(b)  \,,\, \left< \mathcal{A} \right> =  \int F \mathrm{d}^2 x \sim \xi(b)^2 \,.
\end{align}
Both terms have a nontrivial $b$ dependence. So in 1D, the cluster number is
\begin{equation}
    \left< N_c \right> = \partial_b \left< \ell \right>  \sim \xi(b) / b \,,\, (d = 1) \,.
\end{equation}
In 2D, we have 
\begin{equation}
    \left< \mathcal{P} \right> = 2 \partial_b \left< \mathcal{A}\right> \sim  \xi(b)^2 / b \,,\,
       \left< N_c - N_h  \right> = \frac14 \partial_b \left< \mathcal{P} \right> \sim \xi(b)^2 / b^2 \,,\, (d = 2) \,. \label{eq:sandwich}
\end{equation}
Now since $\mathcal{P} / (4b) \ge  N_c \ge N_c - N_h$, the above results sandwich the asymptotics of $\left< N_c \right>$:
\begin{equation}
     \left< N_c   \right> \sim \xi(b)^2 / b^2 \,,\, (d=2) \,.
\end{equation}
This sandwiching argument will be systematically repeated below to obtain the cluster number asymptotics in 2D, see \eqref{eq:Ncnontrivialsmall}, \eqref{eq:plateaugap}, \eqref{eq:gap_linear}, \eqref{eq:plateaugap1} below. 

\subsection{Application to the Covid-19 outbreak in the United States}
As a proof of principle of our method, we test our approach against the real-world on the Covid-19 outbreak in the United States in March 2020. The long-range dispersal is important for describing the epidemic spreading in human society. Indeed, it has been shown~\cite{brockmann,gonza08nature} that human mobility is well described by a L\'evy flight with $\alpha \approx 0.6$, with a cutoff of $\sim 10^3 \text{km}$. 

We test the prediction of our model in the supercritical regime, on the distribution of gaps between clusters. We recall that this is defined as the number of clusters $N_c$ as a function of the coarse-grain distance $b$. The prediction is that its average value is time-independent, up to a normalisation that depends on the number of infections $S$. More precisely, 
\begin{equation}
    \left< N_c(b) \right> \sim S^{\frac{d}{\alpha+d}} b^{-\frac{\alpha d}{\alpha + d}} \,. \label{eq:prediction}
\end{equation}

We test this prediction against the data on the initial outbreak of Covid-19 in the United States (continental states) in March 2020. County-level daily infection numbers are made available by The New York Times (\href{https://github.com/nytimes/covid-19-data}{https://github.com/nytimes/covid-19-data}). For any given day, we obtain a set of points which are the geographical center of the counties where infections have been reported, see Fig.~\ref{fig:US} (top). Then, for any distance $b$, we form a graph by connecting all pairs of infected counties with geodesic distance $\le b$, and compute $\left< N_c (b) \right> $ as the number of connected components of the graph. We can extract the total infection number $S$ for each day. 

The results are shown in Fig.~\ref{fig:US} (bottom). We found that in a time window of roughly a week (March 1 - March 5), and for $10 \text{km} \leq b \leq 10^3 \text{km}$,  $ \left< N_c(b) \right> $ is time-independent up to a global pre-factor that increases with time. Upon dividing by $S^{-\frac{d}{\alpha+d}}$, see \eqref{eq:prediction}, the data for different days are collapsed. Moreover, the $b$ dependence is consistent with a power law $ b^{-\frac{\alpha d}{\alpha + d}}$, where $\alpha \approx 0.6$ as previously found by independent studies~\cite{brockmann,gonza08nature} (one of the works measured $\alpha$ by tracking the displacement of dollar bills). After the first week of March, the epidemic covers almost all the counties of the United States, see Fig.~\ref{fig:US} (top). Thus, the behavior of $ \left< N_c(b) \right> $ changes qualitatively and is no longer described by our theory.

% To compare to our prediction \eqref{eq:prediction}, we use the estimated L\'evy flight exponent for human mobility $\alpha \approx 0.6$. We find an encouraging agreement in Fig.~\ref{fig:US} (bottom). Indeed, after that, $\left< N_c(b) \right>$ slowly converges to another time-independent distribution, not described by our theory. 
%We surmise that this distribution reflects the non-uniform distribution of the counties, and has little to do with the epidemics.

It may seem concerning that the prediction works only for a few days. However, this is consistent with the rapid growth of the outskirt diameter $D$ predicted by our model. Indeed, the daily infection data we used indicate that the total number of infections doubles every two days in the beginning of March 2020, $S \sim 2^{n_{\text{days}} / 2}$. By the scaling law $D \sim S^{1/\alpha}$, the diameter of the epidemic doubles every day. The distance resolution in our analysis is about $10 \text{km}$, and the L\'evy nature of human mobility is valid up to $\sim 10^3 \text{km}$~\cite{brockmann,gonza08nature}, giving us a window of two orders of magnitude. Therefore the number of days where our theory is expected to work is 
$$ n_{\text{days}} = \frac{\ln (10^2)}{\ln(2) / (2\alpha)} \approx 5.5 \,, $$
which is approximately a week.

To summarize, we tested the prediction on the gap distribution against real-world epidemic data and obtained an encouraging agreement in the initial stage of the outbreak. The quantitative prediction~\eqref{eq:prediction} appears to be robust despite many real-world factors that are not taken into account. This is a demonstration of universality in statistical physics. 

\begin{figure}
    \centering
    \includegraphics[width=.95\textwidth]{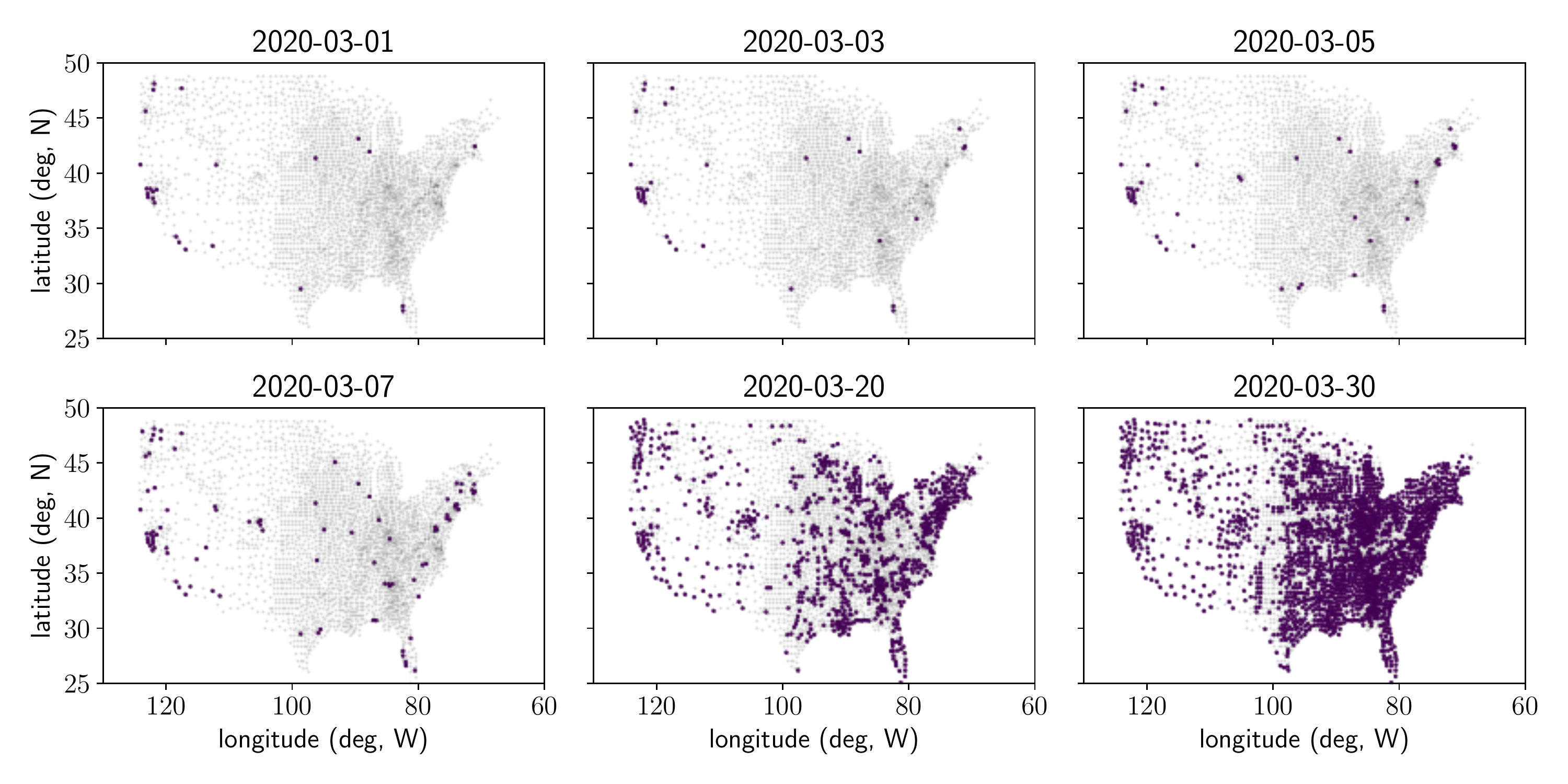}
    \includegraphics[width=.98\textwidth]{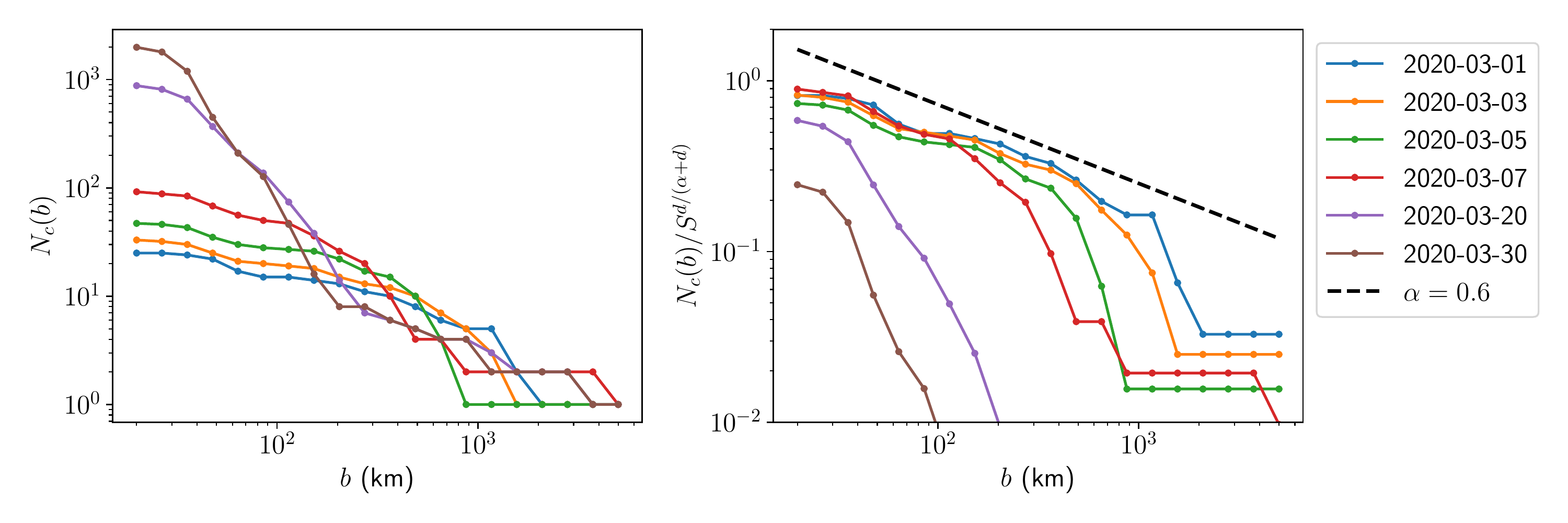}
    \caption{\textit{Top}. The purpose dots are the geographical center of the counties of the continental US where Covid-19 infections have been reported. The grey dots are all other counties. \textit{Bottom}. 
    The number of connected components of the graph obtained by connecting infected counties with distance $\le b$, as a function of $b$, for several days in March 2020. The data in early March is compared to the prediction \eqref{eq:prediction}, with $d=2$ and $\alpha = 0.6$ (dashed line). }
    \label{fig:US}
\end{figure}

\subsection{Critical regime}
\subsubsection{General strategy}
We approach criticality from the sub-critical side, where the epidemic always goes to extinction. Thus we shall always consider the $t\to\infty$ limit. For simplicity, we may let
\begin{equation} \label{eq:subcritical}
 \beta = 1 \,,\, \gamma = 1 + m^2  \,,
\end{equation}
where $m^2$ is a small positive number (``mass squared'' in the avalanche context) that controls the distance to criticality. 

We can obtain averages conditioned on the total infection number $S$ by deriving with respect to $m^2$. Indeed, this is because the distribution of the total infection number, $S$, follows a power law $P(S) \sim S^{-3/2}$ with a cutoff at $S_m = m^{-4}$. Therefore, for any observable $\mathcal{O}$, we have
\begin{equation}
      \left< \mathcal{O} \right> \sim \int^{S_m} S^{-3/2}     \left< \mathcal{O} \right>_S  \mathrm{d} S
\end{equation}
where $\left< \mathcal{O} \right>_S $ is the average at criticality and conditioned on $S$, while $\left< \mathcal{O} \right>$ is the (non-conditioned) average over the near-critical ensemble \eqref{eq:subcritical}. Differentiating both sides with respect to $m^2$, we obtain
\begin{equation} \label{eq:OSm2}
    \left< \mathcal{O} \right>_S \sim - \left. \frac{\partial \left<  \mathcal{O} \right>}{\partial {m^2}} \right\vert_{m^2 \to S^{-\frac12}} \,.
\end{equation}
Most often, $\left< \mathcal{O} \right>$ contains a term that is proportional to a power of $m^2$. Then, $\left< \mathcal{O} \right>_S $ is proportional to that term, multiplied by $m^{-2} \sim \sqrt{S}$.

We can also directly obtain finite time averages at criticality from the $t=\infty$, sub-critical one, by simply substituting $m^2 \to 1/t$. This is because the critical and near-critical dynamics are indistinguishable until $t \sim m^{-2}$, after which the sub-critical one saturates. 

In view of the above considerations, we shall concentrate on the the stationary instanton equation in $d$ dimensions, $d=1, 2$:
\begin{align} \label{eq:stationary-supp}
  \mathcal{D}^\alpha F = F^2 + m^2 F  \,,\, \Vert x \Vert > b/2  \,,\,  
    F( \Vert x \Vert < b/2) = 1 \,.
\end{align}
Indeed, we verified numerically that the time-dependent instanton equation always converges (point-wise) to a stationary solution as $t\to\infty$ in the subcritical regime. The integral of $F$, and its $b$-derivative provides the sub-critical averages. To analyze the asymptotic behavior of the solution, we shall consider a few approximate solutions to it. Each of them is dominant in some range of parameters. Then we show how to assembly them in various regimes. In what follows, we shall assume $\alpha < 2$, until Section~\ref{sec:shortrange}, where the competition with short-range physics is discussed. 

The main results of the analysis below are summarized in Table~\ref{table}.

\textit{Note}. Unless otherwise stated, we assume that $b \ge 1$, and restrain from considering smaller values of $b$. Considering $b \ge 1$ is enough for deriving the results of the main text. We will comment on situations where considering $b \ll 1$ might be useful (see Section~\ref{sec:shortrange} below).

\begin{table*}[]
	\centering
	
	\begin{tabular}{|c|c|c|c|c|c|}\hline
		$d=1$ &  $\alpha < 1/2$ & $1/2 < \alpha < 1$  & $1< \alpha < 3$ & $3< \alpha < 4$ & $\alpha > 4$ \\ \hline
		$ \left< \ell \right>_S \sim S^{?}  $ &
		$1$ & $ {1/(2\alpha)} $ & $ {1/(1+\alpha)} $ & \multicolumn{2}{c|}{${1/4}$}  \\ \hline
		$ \left< N_c \right>_S \sim S^{?} $ &
		$1$ & $ {\chi/(2\alpha)} $ & $ {1/(1+\alpha)} $ & ${1-\alpha/4}$ & $0$ \\ \hline
		$ \left< \ell(t) \right> \sim t^{?} $ &
		$1$ & $ {(1-\alpha)/\alpha} $ &  \multicolumn{3}{c|}{$0$}  \\ \hline
		$ \left< N_c(t) \right> \sim t^{?} $ &
		$1$ & $ {(\chi-\alpha)/\alpha} $ &  \multicolumn{3}{c|}{$0$}  \\ \hline
	\end{tabular}
	\begin{tabular}{|c|c|c|c|c|}\hline
		$d=2$  &  $\alpha < 1$ & $1 < \alpha < 2$  & $2 < \alpha < 4$ & $\alpha> 4$ \\ \hline
		$ \left< \mathcal{A} \right>_S \sim S^?  $ &
		$1$ & $ {1/\alpha} $  & \multicolumn{2}{c|}{${1/2}$}  \\ \hline
		$ \left< N_c \right>_S \sim S^? $ &
		$1$ & $ {\chi/(2\alpha)} $  & ${1-\alpha/4}$ & 1 \\ \hline
		$ \left< \mathcal{A}(t) \right>\sim t^?  $ &
		$1$ & $ {(2-\alpha)/\alpha} $  & \multicolumn{2}{c|}{${0}$}  \\ \hline 
		$ \left< N_c(t) \right>\sim t^?  $ &
		$1$ & $ {(\chi-\alpha)/\alpha} $  & \multicolumn{2}{c|}{${0}$}  \\ \hline
	\end{tabular}
	\caption{Summary of the main asymptotic results in the critical regime. Scaling exponents are displayed for both the time average and conditioned average on a large infection number $S$. In two dimensions, the perimeter's scaling is identical to the cluster number. }
	\label{table}
\end{table*}

\subsubsection{Extent of the bulk and the outskirt}
Before proceeding with the analysis of the instanton equation, we recall the simpler calculation, of the average density $S(x)$ of infected individuals at distance $x$ from the origin [the size of the infected population is is the integral of $S(x)$]. By a similar backward recursion argument as above, we can show that it satisfies a linear equation
\begin{equation}
     \mathcal{D}^\alpha S(x) = m^2 S(x) - \delta(x) \,. 
\end{equation}
Therefore, $S(x)$ is nothing but the Green function of a fractional Gaussian free field with mass $m^2$ and a kinetic term $\propto |k|^\alpha$ when $\alpha < 2$ (and $|k|^2$ when $\alpha > 2$, see \eqref{eq:pSR} below) in the momentum space. The bulk extent is the correlation length of this field: 
\begin{equation}
    \begin{cases} 
\xi = m^{-\frac2\alpha} & \alpha< 2 \\ \xi_{\text{SR}} =  m^{-1}  & \alpha > 2 \end{cases} \,.
\end{equation}
({Here}, in the Supplemental Material, to avoid confusion, we shall use the subscript $\text{SR}$ to denote the short-range bulk extent. The symbol $\xi$ without subscript is always equal to $m^{-2/\alpha}$ even when $\alpha > 2$.) Beyond the correlation length, $S(x) \sim |x|^{-\alpha-d}$ has a fast-decaying tail. Therefore most of the infections happen inside the bulk. Identifying $m^2 = S^{-1/2}$ according to \eqref{eq:OSm2} gives the expressions in the main text: $\xi = S^{1/(2\alpha)}$ for $\alpha <2$ and $\xi_{\text{SR}} = S^{1/4}$ for $\alpha > 2$. 

We also recall the simple argument leading to the scaling law of the outskirt radius $D \sim S^{1/\alpha}$, Eq. (3) of the main text. Indeed, consider $S$ independent jump distances $r_1, \dots, r_S$, each distributed according to Eq (1) of the main text. It is not hard to see that $\mathrm{Prob}(|r_i| < x) \sim  1 - C  x^{-\alpha}$ for any $i$. By independence, $\mathrm{Prob}(\max(|r_1|, \dots |r_S|) < x) \sim (1-C x^{-\alpha})^{S} \sim \exp(-C x^{-\alpha } S) $. Thus, the typical value of the maximal jump size is $r_{\max }\max(|r_1|, \dots |r_S|) \sim S^{1/\alpha} $. Now, assuming that the radius of the outskirt is dominated by $r_{\max}$, we obtain Eq. (3) of the main text. Although this argument seems heuristic, Eq. (3) is exact, as confirmed by the systematic analysis below, see remarks around Eq.~\eqref{eq:nearplateau}.

\subsubsection{Scale invariant solution and subleading term}
We now come back to the analysis of the {time-independent} instanton equation~\eqref{eq:stationary-supp} and consider its first approximate solution. The ``scale invariant'' approximation ignores the mass term and the boundary condition at $\Vert x \Vert < b/2$, and focuses on power law type solutions of the equation
\begin{equation} \label{eq:massless}
     F^2 = \mathcal{D}^\alpha F \,.
\end{equation}
The mass introduces a cutoff of this equation when $F \sim m^2$ (since that is where $m^2 F \sim F^2$ in \eqref{eq:stationary-supp}).

To solve this equation we recall the classic formula on the Fourier transform of power laws:
\begin{equation}\label{eq:powerlawFT}
    \mathcal{F}[|x|^{-a}] := \int |x|^{-a} e^{\im x.k} \mathrm{d}^d x = B(a) |k|^{a-d} \,, \text{where } B(a) = \frac{\pi ^{d/2} 2^{d-a} \Gamma \left(\frac{d-a}{2}\right)}{\Gamma \left(\frac{a}{2}\right)} \,.
\end{equation}
Note that $B$ also depends on $d$ but we omitted this argument to keep notations concise. Then, for $\alpha < 2$, the fractional diffusion term $\mathcal{D}^\alpha$ acts on a power law in the following way:
\begin{align}
 &   \mathcal{D}^\alpha (|x|^{-h}) = |x|^{-h-\alpha} D(h, \alpha) \,, \label{eq:Dxa}  \\
 &   D(h,\alpha) := \frac{1}{(2\pi)^d}  B(h) B(d-h-\alpha) B(d+\alpha)  \,. \end{align}
This can be shown by performing the convolution by Fourier transform: $\mathcal{D}^\alpha f = \mathcal{F}^{-1} \left( \mathcal{F}[p]\mathcal{F}[f]\right)$. 

From \eqref{eq:Dxa}, it follows immediately that \eqref{eq:massless} admits a power law solution, which we shall call the scale-invariant approximation:
\begin{equation}
     F_{\text{sc}}= D(\alpha,\alpha) |x|^{-\alpha}  \,. \label{eq:Fsc-supp}
\end{equation} 
Comparing this to $m^2$, we find the mass cutoff
\begin{equation} 
     \xi = m^{-2/\alpha} \,,
\end{equation}
which is the same as the bulk extent. When $|x|\gg \xi$, $F_{\text{sc}} \sim |x|^{-\alpha-d}$ decays fast and its contribution can be ignored for all purposes. Plotting $D(\alpha,\alpha)$, one may find that it is positive when $\alpha \in (d/2,d)$. It has a zero at $\alpha = d/2$ and a pole at $\alpha = d$. Therefore the scale invariant solution is valid as a dominant asymptotic behavior \textit{only} in the interval $\alpha \in (d/2,d)$. A similar solution was found in 1D in \cite{lepriol-thesis}. 

\textit{Remark.} We have swept some (well-known) technical details under the rug. Indeed, $\mathcal{F}[p]$ is the regularized Fourier transform, with an $\epsilon$-dependent constant removed, and to which the formula \eqref{eq:powerlawFT} applies. Also, the RHS of \eqref{eq:Dxa} also misses $\delta$ terms at the origin, which are unimportant since our analysis concerns large $x$.  

\subsubsection{Correction to scale invariant solution}
It will be important to consider admissible perturbations of the scale invariant solution $F_{\text{sc}}$. That is, we consider 
\begin{equation}
    F = F_{\text{sc}} + \delta F \,,\, \delta F \ll F_{\text{sc}} \,.
\end{equation}
Such an $F$ satisfies \eqref{eq:massless} if $\delta F$ satisfies its linearized version
\begin{equation}
    \mathcal{D}^\alpha (\delta F) = 2 F_{\text{sc}} \delta F \,.
\end{equation}
This also admits a power-law solution at large distances
\begin{equation}\label{eq:eta}
     \delta F \propto  |x|^{-\eta} \,, \text{ where $\eta$ satisfies }
    2 D(\alpha, \alpha)= D(\eta,\alpha)  \,.
\end{equation}
Anticipating what follows, we note that the exponent $\eta$ is related to $\chi$ in the main text by
\begin{equation}
    \chi = d + \alpha - \eta \,. \label{eq:kappaeta}
\end{equation}
See Fig.~\ref{fig:kappa12} for a plot of $\chi$. The explicit form of the  transcendental equation $ 2 D(\alpha, \alpha)= D(\eta,\alpha)$~\eqref{eq:eta}, in terms of $\chi$, is the following:
\begin{equation}\label{eq:manygamma}
  \frac{D(\eta,\alpha)}{2D(\alpha, \alpha)} =  \frac{\Gamma \left(\frac{\alpha }{2}\right) \Gamma \left(\frac{\chi -\alpha }{2}\right) \Gamma \left(\frac{1}{2} (d-2 \alpha )\right) \Gamma \left(\frac{1}{2} (d+2 \alpha -\chi )\right)}{2 \Gamma (\alpha ) \Gamma \left(\frac{1}{2} (\chi -2 \alpha )\right) \Gamma \left(\frac{d-\alpha }{2}\right) \Gamma \left(\frac{1}{2} (d+\alpha -\chi )\right)} = 1 \,.
\end{equation}
{This equation determines $\chi$ as mentioned in the main text.} {Note that for each $d,\alpha$ there are {\it several branches of solutions} to this equation (generically an infinity, but only two for $\alpha=2,4,..$).} The branch relevant to this study is the unique one such that $\chi \in (\alpha, d)$ (or $\eta \in (\alpha, d)$) as $\alpha \in (d/2, d)$. We justify this choice by observing that $\chi \to d$ as $\alpha \to d/2$, which is expected from the continuity to the $\alpha < d/2$ regime (see below).
\\

{\it Remark}. A similar analysis of the linear perturbation of the instanton equation
around the self-consistent solution $F_{\text{SR}}(x) = 2(4-d)/x^2$ was performed for the short-range Brownian 
model (in the continuum setting of the Brownian force model) \cite{pierrenewBFM} (Section VIII) 
and led to two possible values for the exponent
$\eta=3,-4$ in $d=1$, $\eta=\pm 2\sqrt{2}$ in $d=2$ and $\eta=\frac{1}{2} \pm \frac{\sqrt{17}}{2}$
in $d=3$. Similar exponents also appeared in calculations of the fractal dimension of the
boundary of the super-Brownian motion \cite{MytnikBoundary}. 

One can ask how the equation \eqref{eq:manygamma} recovers the SR case. This happens by setting
$\alpha=2$, while keeping $d, \eta$ generic. In that case \eqref{eq:manygamma} simplifies and one obtains two branches of solutions:
\be  \label{eq:SReta}
\frac{D(\eta,2)}{2D(2,2)} = \frac{\eta  (d-\eta -2)}{4 (d-4)}
\quad \Rightarrow \quad 
\eta = \frac{1}{2} \left(d \pm \sqrt{(d-20)  d+68}-2\right) \,,
\ee
which recover the above cited values.

It is interesting to note however that the $\alpha \to 2$ limit of the transcendental equation~\eqref{eq:manygamma} is quite subtle in 2D. First of all, the $\alpha \to2$ and $d\to 2$ limits do not commute. We have the following series expansion of the LHS of \eqref{eq:manygamma}:
\begin{equation}\label{eq:dalpha}
    \left.\frac{D(\eta,\alpha)}{2D(\alpha, \alpha)}\right|_{\substack{d = 2-\epsilon \\ \alpha = 2 - r \epsilon}} = \frac{\eta^2 (r -1)}{16 r -8} + \mathrm{O}(1) \,,\, {\epsilon \to 0}. 
\end{equation}
The short-range model is obtained by taking $r = 0$, i.e., sending $\alpha\to2$ \textit{before} $d\to 2$ as noted above (indeed, equating the above equation to $1$, we obtain $\eta = \pm 2\sqrt{2}$ in agreement with above). On the other hand, taking $r = \infty$, i.e., sending $d\to2$ before $\alpha \to 2$, we obtain $\eta \to \pm 4$ as $\alpha \to 2$. By tuning $r$, i.e. the angle of approach to $(\alpha,d)= (2,2)$, we can obtain any other value of $\eta$ as a limit. However, the solution that is useful for the cluster statistics in this work is a different branch, and invisible from the series expansion~\eqref{eq:dalpha}. This is because at $d = 2$, the limits $\eta\to2$ and $\alpha \to 2$ do not commute either. We have 
\begin{equation}
      \left.\frac{D(\eta,\alpha)}{2D(\alpha, \alpha)}\right|_{\substack{d = 2 \\ \alpha = 2 -  \epsilon \\ \eta =2 - r \epsilon}} = \frac{1+r}{4r} + \mathrm{O}(1) \,,\, {\epsilon \to 0}. 
\end{equation}
Equating the RHS to $1$ gives $r = 1/3$, or $2-\eta = (2-\alpha)/3 + \mathcal{O}((2-\alpha)^2)$ at $d=2$. We note in passing that in 1D, $1-\eta = (1-\alpha)/2 + \mathcal{O}((1-\alpha)^2)$. 

\begin{figure}
    \centering
    \includegraphics[width=.7\textwidth]{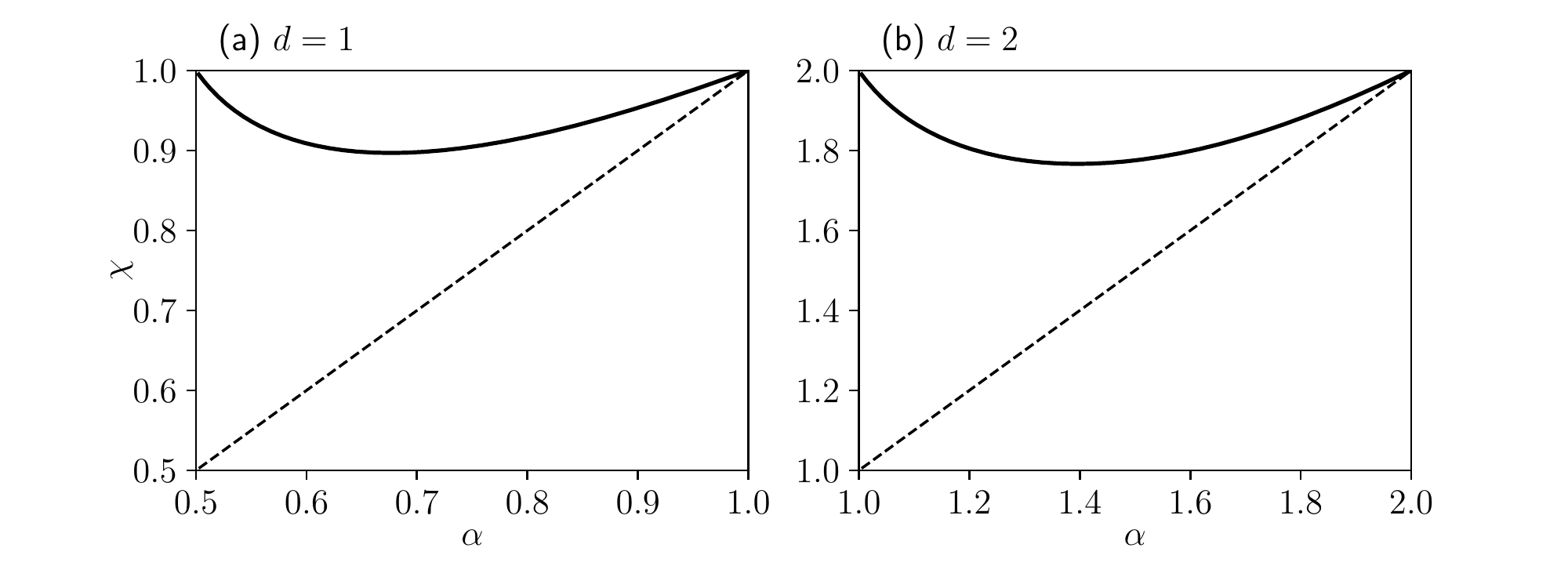}
    \caption{Solid curve: The exponent $\chi$ as a function of $\alpha$ for $d=1$ and $d=2$. The dashed line is $\alpha$. }
    \label{fig:kappa12}
\end{figure}

\subsubsection{The plateau approximation}
We now consider another approximate solution: the ``plateau approximation''. It consists in replacing the LHS of \eqref{eq:stationary-supp} by $\mathcal{D}^\alpha F  \approx  \mathcal{D}^{\alpha}\left[\theta(b/2-\Vert x \Vert)\right]$, which is calculated in \eqref{eq:Dplateau-sup}. Equating that to $F^2$, we obtain
\begin{equation}\label{eq:Fpl-supp}
    F_{\mathrm{pl}} =  b^{\frac{d}2}  |x|^{-\frac{\alpha + d}2} \,,\,  b \ll |x| \ll X_m
\end{equation}
where the mass cutoff scale can be again determined by $ F_{\mathrm{pl}}(X_m) = m^2$:
\begin{equation}
    X_m= b^{\frac{d}{\alpha+d}} m^{-\frac{4}{d+\alpha}} \,. \label{eq:Xm}
\end{equation} 
Beyond that $F_{\mathrm{pl}} \sim |x|^{-\alpha -d}$ decays fast and its contribution can be neglected.

Under the plateau approximation,  the area or the extension is given by
\begin{equation}
    \int F_{\text{pl}}(x) \mathrm{d}^d x = \begin{dcases}   c_0(b) - c_1 X_m^{d} m^2 & \alpha > d \\
      c_1  X_m^{d} m^2  &  \alpha < d
    \end{dcases} 
\end{equation}
where $c_1$ is some positive constant independent of $b$, while $c_0(b)$ depends on $b$ but not on $m^2$. When $\alpha < d$, the above formula leads to the large gap power law of the gap distribution, see \eqref{eq:plateaugap} below. When $\alpha>d$, the integral of $F_{\text{pl}}$ remains of order one 
as $m^2 \to 0$. Yet, the subleading term $ c_1 X_m^{d} m^2$ is $m^2$ dependent, and can be used to calculate the conditioned average on a large $S$, via~\eqref{eq:OSm2}:
\begin{equation}
   \partial_{m^2}  \int F_{\text{pl}}(x) \mathrm{d}^d x  \sim X_m^{d} \,,\, \alpha > d \,. \label{eq:plateauS}
\end{equation}

\textit{Remark.} When the point is near the $b$-neighborhood, $\Vert x \Vert - b/2 \ll b$, the plateau approximation has a slower decay:
\begin{equation}
    F_{\mathrm{pl}} \sim  |\Vert x \Vert - b/2|^{-\frac{\alpha}2} \,,\, 0 <  \Vert x \Vert - b/2 \ll b \label{eq:nearplateau}
\end{equation}
However, this regime is important only if $X_m \lesssim b$, which is equivalent to $b \gtrsim D = m^{-4/\alpha}$. To make the following discussion less cumbersome, we shall always assume $b \ll D$ and ignore the near-plateau regime~\eqref{eq:nearplateau}, unless otherwise stated. (Larger values of $b$ no longer probe the gap distribution but rare instances of gaps greater than $D$.)

\subsubsection{Solution for $d/2 < \alpha < d$}
We are now ready to build the solution for the most interesting regime $d/2 < \alpha < d$, using the above pieces. Since $\alpha < d$, $(\alpha+d)/2>\alpha$, we expect that the solution is dominated by the plateau approximation~\eqref{eq:Fpl-supp} at small distances and by the scale invariant one~\eqref{eq:Fsc-supp} at large distances. Comparing them we obtain a crossover scale 
\begin{equation}\label{eq:xb}
    F_{\text{pl}}(x_b) =  F_{\text{sc}}(x_b) \Rightarrow x_b = b^{\frac{d}{d-\alpha}} \,.
\end{equation}
However, recall that the scale-invariant solution has a cutoff at $|x| \sim \xi$, and can exist if and only if $x_b \ll \xi$. Otherwise, the plateau approximation dominates everywhere, up to a larger mass cutoff $X_m \gtrsim \xi$. The crossover value of $b$ is the crossover gap scale 
\begin{equation}
    x_b \sim \xi \Rightarrow b \sim g_c = \xi^{1-\alpha/d} \label{eq:gc-supp}
\end{equation}
we referred to in the main text. In summary we have:
 \begin{align} \label{eq:F-interesting}
       & F = \begin{dcases}
        F_{\text{pl}} & |x| \ll x_b  \\
        F_{\text{sc}} + \delta F & x_b \ll |x| \ll \xi
        \end{dcases} \,,\,\text{if } b\ll g_c  \\
        &F =  F_{\text{pl}} \,,\, |x| \ll X_m  \,,\, \text{if } b\gg g_c \,. \label{eq:Fplbgc}
\end{align} 
Above, $\delta F$ is the subleading correction to the scale invariant solution. We have seen that it must be proportional to $|x|^{-\eta}$~\eqref{eq:eta}. It remains to fix the prefactor, which will turn out to be $b$-dependent. To do this, we argue that a separation of the scales $b\ll x_b \ll \xi$ imposes a single-parameter scaling of $F$ near $x_b$:
\begin{equation} \label{eq:scalingAnsatz}
    F(x\vert b) = x_b^{-\alpha} \tilde{F}(x/x_b) \,,
\end{equation}
for some scaling function $\tilde{F}(y)$ such that $\tilde{F}(y\to \infty) \sim y^{-\alpha}$ and $\tilde{F}(y\to0) \sim y^{-(\alpha+d)/2}.$ This scaling form is fixed by the leading terms in \eqref{eq:F-interesting}. We verified this ansatz with extensive numerical solution of the instanton equation, see Fig.~\ref{fig:collapse}. Imposing this scaling form to the subleading term fixes its prefactor 
\begin{equation}
    \delta F \sim |x/x_b|^{-\eta} x_b^{-\alpha}  \sim  b^{\frac{d(\eta-\alpha)}{d-\alpha}} |x|^{-\eta}  \,. \label{eq:deltaF}
\end{equation}
It indeed has a nontrivial $b$-dependence, which will allow us to obtain the number of clusters and the bulk gaps distribution, see below.

We now have the complete solution in the regime $\alpha \in (d/2,d)$ (we keep all the dependence on $x$, $m^2$ and $b$, but drop out all other prefactors):
\begin{empheq}[box=\widefbox]{align}
         & F \sim \begin{dcases}
         F_{\text{pl}} = b^{\frac{d}2} x^{-\frac{\alpha+d}2} & |x| \ll x_b = b^{\frac{d}{d-\alpha}} \\
         F_{\text{sc}} + \delta F =  |x|^{-\alpha} +  x_b^{\eta - \alpha} |x|^{-\eta}  & x_b \ll |x| \ll \xi = m^{-2/\alpha}
        \end{dcases} \,,\,\text{if } b\ll g_c = \xi^{1-\alpha/d} \nonumber  \\
        &F \sim     F_{\text{pl}} =  b^{\frac{d}2} x^{-\frac{\alpha+d}2} \,,\, |x| \ll X_m = b^{\frac{d}{\alpha+d}} m^{-\frac{4}{d+\alpha}} \,,\, \text{if } b\gg g_c \,. 
        \label{eq:Fnontrivial}
\end{empheq}
When $|x| \gg \xi$ or $|x|\gg X_m$, $F$ decays as $|x|^{-\alpha-d}$ and can be ignored.

To find the mean area and extension, we take $b \ll g_c$, and integrate over $F$. From~\eqref{eq:Fnontrivial} we can see that the integral is dominated by $F_{\text{sc}} +\delta F$ at its mass cutoff scale $|x|\sim \xi$, so that:
\begin{equation}
    \left< \ell(b) \right>, \left< \mathcal{A}(b) \right> \sim \int_{|x|<\xi} F(x) \mathrm{d}^d x \sim \xi^{d-\alpha} + x_b^{\eta-\alpha} \xi^{d-\eta} \,.
\end{equation}
(Here and below, the RHS with $d=1$ applies to $\ell$, and $d=2$ applies to $\mathcal{A}$.) Note that the first term in the RHS is dominant but $b$-independent; only the subdominant one is $b$-dependent. Applying \eqref{eq:OSm2}, setting $b=1$ and neglecting the subdominant term, we obtain the $S$-conditioned mean extension/area of the main text. 

The mean cluster number and the bulk gap $(b\ll g_c)$ distribution is dominated by the subleading term of $F$ involving $\eta$, $|x| \sim \xi$, since the leading term is $b$-independent. The result can be written in a nice way using $g_c$:
\begin{equation}
    \left< N_c(b=g) \right> \sim \partial_b^d \int_{|x|<\xi} F \mathrm{d}^d x  \sim (g/g_c)^{-\frac{d(d-\eta)}{d-\alpha}} = (g/g_c)^{-\frac{d(\chi -\alpha)}{d-\alpha}} \,,\, g \ll g_c \,. \label{eq:Ncnontrivialsmall}
\end{equation}
Here we recall that  $\chi = \alpha + d - \eta$ by definition. Upon applying \eqref{eq:OSm2} we find the small-gap result and the cluster number (with $b=1$) result of the main text. We also recall that in 2D, we need to use the sandwiching argument $\left< \mathcal{P} \right> \gtrsim \left< N_c \right> \ge \left< N_c - N_h \right>$, as discussed below \eqref{eq:sandwich}.

Similarly, the outskirt gap distribution is dominated by the $b \gg g_c$ case of \eqref{eq:Fnontrivial}. We have 
\begin{equation}
    \left< N_c(b = g) \right> \sim \partial_b^d \int_{|x|<X_m} F \mathrm{d}^d x \sim  (g/g_c)^{\frac{\alpha d}{\alpha + d}}  \,,\, g \gg g_c \,.\label{eq:plateaugap}
\end{equation}
The $S$-conditioned result in the main text is then found by applying \eqref{eq:OSm2}. Note that, we can see clearly from the calculation above that these gaps are in the outskirt, by noticing that the integral of $F$ is dominated by the scale $X_m$, which is larger than $\xi$.

\begin{figure}
    \centering
    \includegraphics[width=.6\textwidth]{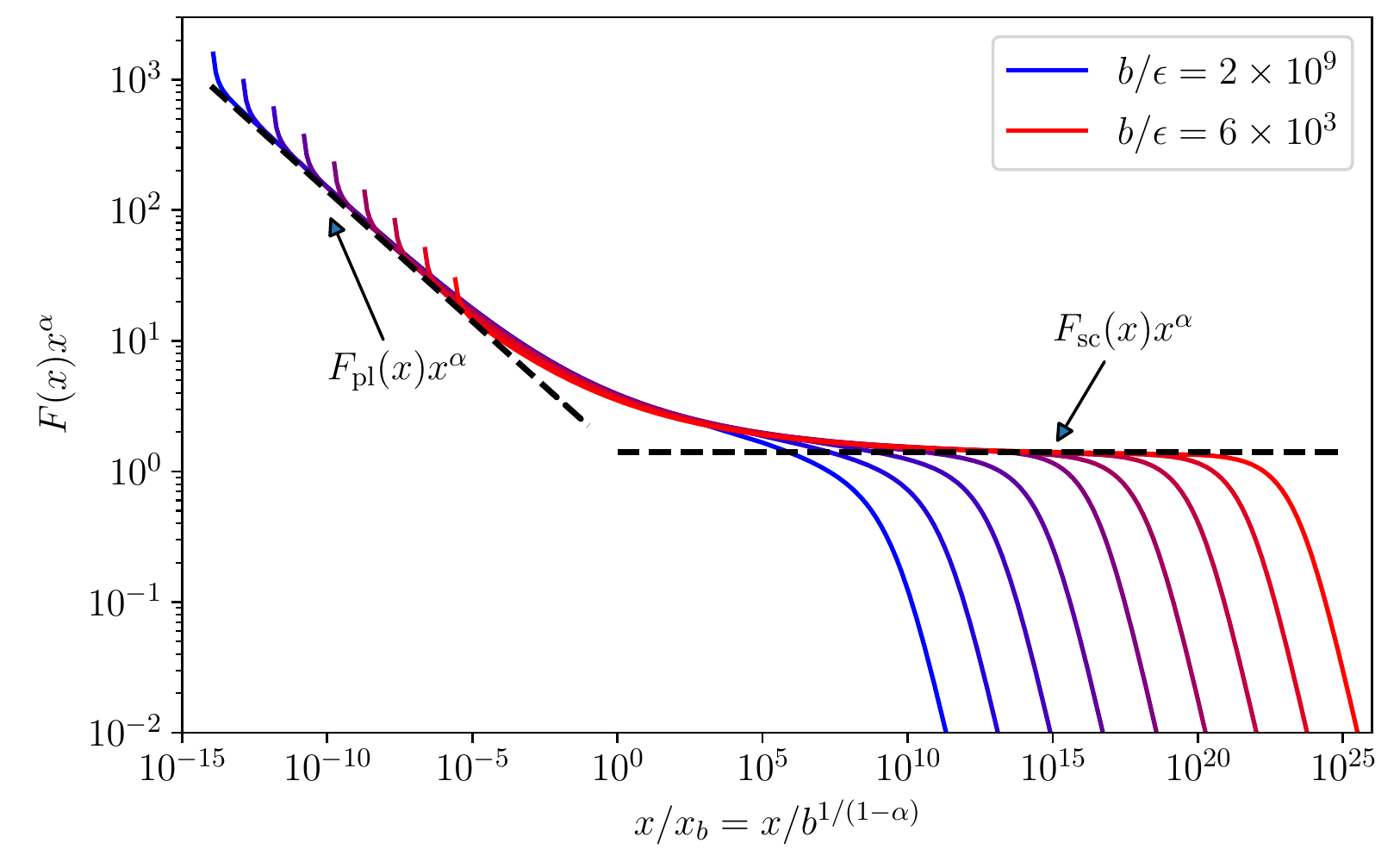}
    \caption{Verification of the scaling ansatz~\eqref{eq:scalingAnsatz}, which is equivalent to $F(x\vert b) = \hat{F}(x/x_b) x^{-\alpha}$ such that $\hat{F}(y\to\infty) = D(\alpha, \alpha)$ and $\hat{F}(y\to0) \sim y^{(\alpha-d)/2}$. We solved the instanton equation numerically at $m^2 = 10^{-19}$ and for $b/\epsilon=6\times 10^3, \dots, 2\times 10^9$ (from red to blue), and plotted $F x^\alpha$ against $x / x_b$ (for $x > b$); the data collapse near $x \sim x_b$ confirms the scaling ansatz. For $x \gg x_b$ and $x \ll x_b$ the solutions are in good agreement with the scale-invariant ($F_{\text{sc}}$) and plateau ($F_{\text{pl}}$) approximations, respectively. Note that the dashed lines have the exact prefactors; no fit is performed. The large distance and small distance deviations from the collapse are due to the mass cutoff and the near plateau behavior~\eqref{eq:nearplateau}, respectively. }
    \label{fig:collapse}
\end{figure}

In Fig.~3 of the main text, we verified the gap distribution prediction by extensive numerical solution of the instanton equation. The derivative with respect to $b$ an $m^2$ are evaluated numerically as finite differences. 

\subsubsection{$\alpha < d/2$: The linear approximation}
When $\alpha < d/2$, the scale invariant solution is no longer viable in the long distance because its prefactor would be negative. In fact, at large distances, the nonlinearity becomes irrelevant and the linear approximation applies, $ F \approx F_{\text{lin}}$ where $F_{\text{lin}}$ is such that 
\begin{equation}\label{eq:Flin-supp}
     (\mathcal{D}^\alpha - m^2) F_{\text{lin}} = \text{fast decaying term} \implies  F_{\text{lin}} \sim 
    c(b) |x|^{-(d-\alpha)}  \,,\, |x| \ll \xi = m^{-\frac{2}{\alpha}} 
\end{equation}
(Recall that we assume $\alpha < 2$ here). Beyond $\xi$, the solution again decays as $|x|^{-d-\alpha}$ and can be neglected. To fix the $b$-dependent prefactor, we can exploit the crossover to the plateau solution at the mass cutoff scale. A smooth crossover requires that $ F_{\text{lin}}(X_m) \sim F_{\text{pl}}(X_m)$ when $ X_m \sim \xi $. This imposes:
\begin{equation}
    c(b) \sim b^{\frac{d(d-2\alpha)}{d-\alpha}} \,,
\end{equation}
and in turn fixes the crossover scale between the linear and plateau regimes: it is still $x_b = b^{\frac{d}{d-\alpha}}$ as in the regime $\alpha \in (d/2,d)$. So the crossover with the plateau approximation works exactly as in the regime $\alpha \in (d/2,d)$. In summary, the instanton solution for the $\alpha < d/2$ is as follows: 
\begin{empheq}[box=\widefbox]{align}
         & F \sim \begin{dcases}
        F_{\text{pl}} = b^{\frac{d}2} x^{-\frac{\alpha+d}2}  & |x| \ll x_b =  b^{\frac{d}{d-\alpha}}   \\
        F_{\text{lin}} = x_b^{-\alpha} |x/x_b|^{\alpha-d}  & x_b \ll |x| \ll \xi
        \end{dcases} \,,\,\text{if } b\ll g_c = \xi^{1-\alpha/d} \nonumber \\
        &F \sim  F_{\text{pl}} = b^{\frac{d}2} x^{-\frac{\alpha+d}2} \,,\, |x| \ll X_m = b^{\frac{d}{\alpha+d}} m^{-\frac{4}{d+\alpha}}  \,,\, \text{if } b\gg g_c \,. \label{eq:Flinallregime}
\end{empheq}
 The extension and area (for $b\ll g_c$) are obtained by integrating $F$, and are dominated by $F_{\text{lin}}$:
\begin{equation}
 \left< \ell(b) \right>, \left< \mathcal{A}(b) \right> =  \int F(x) \mathrm{d}^d x \sim \xi^{\alpha} x_b^{d-2\alpha} \,,\, b \ll g_c \,.
\end{equation}
Note that these quantities have a nontrivial $b$-dependence already at leading order, at variance with the case  of $\alpha\in (d/2,d).$ Applying \eqref{eq:OSm2} we obtain 
\begin{equation}
    \left< \ell(b) \right>_S, \left< \mathcal{A}(b) \right>_S  \sim S   x_b^{d-2\alpha}  \,.
\end{equation}
Applying the $\partial_b^d$ derivative we obtain the gap distribution and the cluster number; applying \eqref{eq:OSm2} gives us the $S$-conditioned average:
\begin{equation}\label{eq:gap_linear}
    \left< N_c(b=g)\right> \sim  \frac{1}{\sqrt{S}} \left< N_c(b=g)\right>_S \sim \begin{dcases}
     (g/g_c)^{\frac{\alpha d}{d-\alpha}}  \,,\, g \ll g_c \\
       (g/g_c)^{\frac{\alpha d}{d+ \alpha}}  \,,\, g \gg g_c
    \end{dcases}
\end{equation}
In particular we find $\left< N_c \right>_S \sim \xi^{2\alpha} \sim S$. In other words, the number of clusters is proportional to the total infected population. So the size of each cluster is of order unity in average: the clusters are atomic. 

\textit{Remark.} We may understand the $\alpha < d/2$ regime as a degeneration of the $\alpha \in (d/2,d)$ one, in the following sense. The linear approximation is the continuation of the subleading term in the scale-invariant solution, whereas the leading term vanishes. Thereby, the gap distribution and cluster number in the regime $\alpha < d/2$ can be expressed using the same formulas of the regime $\alpha \in (d/2,d)$, upon the following replacements of the exponents:
\begin{equation}
    \eta \to d - \alpha \,,\, \chi \to 2 \alpha \,,\, (\alpha < d/2) \,. 
\end{equation}

\subsubsection{Regimes with $\alpha > d$}\label{sec:shortrange}
When $\alpha > d$, among the above approximate solutions, only the plateau one can survive. However, when $\alpha > 2$, the fractional diffusion operator contains a normal diffusion term: more precisely, in the Fourier space
\begin{equation}
    \mathcal{F} [p](k) = A_\epsilon |k|^2  + \dots + B(\alpha + d) |k|^{\alpha} \,, \label{eq:pSR}
\end{equation}
 where $A_\epsilon$ is a constant depending on the short-distance cutoff of $p_\alpha(x)$, and $\dots$ denotes further even powers ($\le \alpha$) of $|k|$ that may appear.  This implies a new approximate solution: the short-range (SR) scale invariant solution, 
\begin{equation}
    F_{\text{SR}} =  2(4-d) A_\epsilon |x|^{-2} \,,\, |x| \ll \xi_{\text{SR}} = m^{-1} \,.
\end{equation}
Comparing this with the plateau approximation both near the plateau $\sim |x|^{-\alpha/2}$ ~\eqref{eq:nearplateau}, and faraway $\sim |x|^{-(\alpha+d)/2}$~\eqref{eq:Fpl-supp}, we identify two threshold values: $\alpha = 4-d$, and $\alpha = 4$. Hence, there are three cases:
\begin{enumerate}
  \item $\alpha \in (1, 3), d= 1$. Here, the plateau approximation is valid everywhere up to the cutoff $X_m$:
\begin{empheq}[box=\widefbox]{equation}
      F \sim F_{\text{pl}} = b^{\frac{d}2} x^{-\frac{\alpha+d}2} \,,\, |x|\ll X_m  =  b^{\frac{d}{\alpha+d}} m^{-\frac{4}{d+\alpha}} \label{eq:Fplregime13}
  \end{empheq}
 Then the extension for any $b$ has a nontrivial $b$ dependence, as follows:
    \begin{equation}
     \left< \ell(b) \right>_S \sim  \partial_{m^2} \int F(x) \mathrm{d} x = b^{\frac{1}{1+\alpha}} S^{\frac1{1+\alpha}} \,.
    \end{equation}
The gap distribution has only the large-gap regime (again given by the plateau approximation):
    \begin{equation}
      \left< N_c(b = g) \right>_S \sim S^{\frac1{1+\alpha}} g^{-\frac{\alpha}{1+\alpha}}  \sim  \sqrt{S}  (g/g_c)^{-\frac{\alpha  }{\alpha + 1}} \,,\, g_c = \xi^{1-\alpha} \ll 1\,. \label{eq:plateaugap1}
\end{equation}
The above results may appear simple formally, but their physical interpretation is rather subtle. Indeed, the average extension is always much greater than the bulk extent, which is $\xi \sim S^{1/(2\alpha)}$ for $\alpha < 2$, and $\xi_{\text{SR}} \sim S^{1/4}$ for $\alpha > 2$. This transition from a long-range bulk to a short-range one is \textit{invisible} from the asymptotic behavior of $\left< \ell \right>$. This is because the extension is not dominated by the bulk (as is the case for all the other regimes $\alpha < d$), but by the many clusters of the outskirt, at a distance $X_m\gg \xi$ from the origin. Now, concerning the bulk itself, the above results provide only indirect information. For example, we may infer that the bulk should be compact, and devoid of gaps of size $\ge 1$, because the gap distribution does not have a small-gap regime: indeed, in \eqref{eq:plateaugap1} $g_c \ll 1$ for $\alpha > 1$. If is possible for very small $b$, $ b \lesssim g_c \ll 1$, the asymptotic behavior of the solution might be different from \eqref{eq:Fplregime13} above, and reveal further information about the bulk in the regime $\alpha \in (1,3), d=1$. We leave this to future study.

\item $\alpha \in (4-d, 4)$. This is a more tricky case, since there is again a crossover from the plateau to the SR scale invariant approximation. The crossover scales can be worked out in a similar way as in the regime $\alpha\in(d/2,d)$, but their values are different:
\begin{empheq}[box=\widefbox]{align}
   & F = \begin{dcases}
      F_{\text{pl}} = b^{\frac{d}2} x^{-\frac{\alpha+d}2} & |x| \ll b^{\frac{d}{\alpha + d - 4}} \\
      F_{\text{SR}} \sim |x|^{-2}  &  b^{\frac{d}{\alpha + d - 4}} \ll |x| \ll \xi_{\text{SR}} = m^{-1} 
    \end{dcases} \,,\, \text{ if } b \ll g_c' = \xi_{\text{SR}}^{\frac{\alpha-4+d}{d}} \\
    & F =F_{\text{pl}} = b^{\frac{d}2} x^{-\frac{\alpha+d}2} \,,\, |x|\ll X_m  =  b^{\frac{d}{\alpha+d}}  m^{-\frac{4}{d+\alpha}} \,,\,  \text{ if } b \gg g_c'   \,.
\end{empheq}
(We do not need to calculate any correction to $F_{\text{SR}}$, as we explain below.) As a result, the extension and area are dominated by the short-range bulk: \eqref{eq:SRellA} still holds, as in the regime $\alpha > 4$. However, in difference with the latter regime, there are many clusters and gaps in the outskirt. Indeed, for $b  \gg g_c'$, we have the gap distribution given by the plateau approximation:
\begin{equation}
      \left< N_c(b = g) \right>_S\sim  \sqrt{S}  (g/g_c)^{-\frac{\alpha d }{\alpha + d}} \,,\, g_c = \xi^{1-\alpha/d} \,,\, b \gg g_c' \,.
\end{equation}
(Note that $\xi = m^{-2/\alpha} \sim S^{1/(2\alpha)}$ by definition.) As $b$ further decreases below $g'_c$, the short-range solution dominates the mass cutoff scale. So we expect that there are no more gaps in the bulk, and therefore the cluster number no longer grows:
\begin{equation}
     \left< N_c(b \ll g_c') \right>_S \sim \left< N_c(b = g_c') \right>_S \sim \xi_{\text{SR}}^{4 - \alpha}  \sim  S^{1-\alpha/4} \,, \label{eq:Ncalphalarge}
\end{equation} 
in both dimensions. In particular, the cluster number scales as $ \left< N_c(b =1) \right>_S \sim S^{1-\alpha/4}$. The above predictions are tested numerically in 1D, see \eqref{fig:SR}. %Setting $b=1$ we find the cluster number result in the main text.
 
We note that, in the numerical solution of the instanton equation in 1D, we did \textit{not} observe the short-range subleading exponents \eqref{eq:SReta}, which is $\propto |x|^{-3}$ in 1D. Indeed a $c(b) |x|^{-3}$ correction to $F_{\text{SR}}$ would imply that $\left< N_c(b=1) \right>_S \sim \mathcal{O}(1)$, which is inconsistent with the results at large gaps $b \gg g_c'$. This does not contradict our analysis of admissible perturbations of the scale invariant solutions. An admissible perturbation may not necessarily appear.  

\begin{figure}
    \centering
    \includegraphics[width=.65\textwidth]{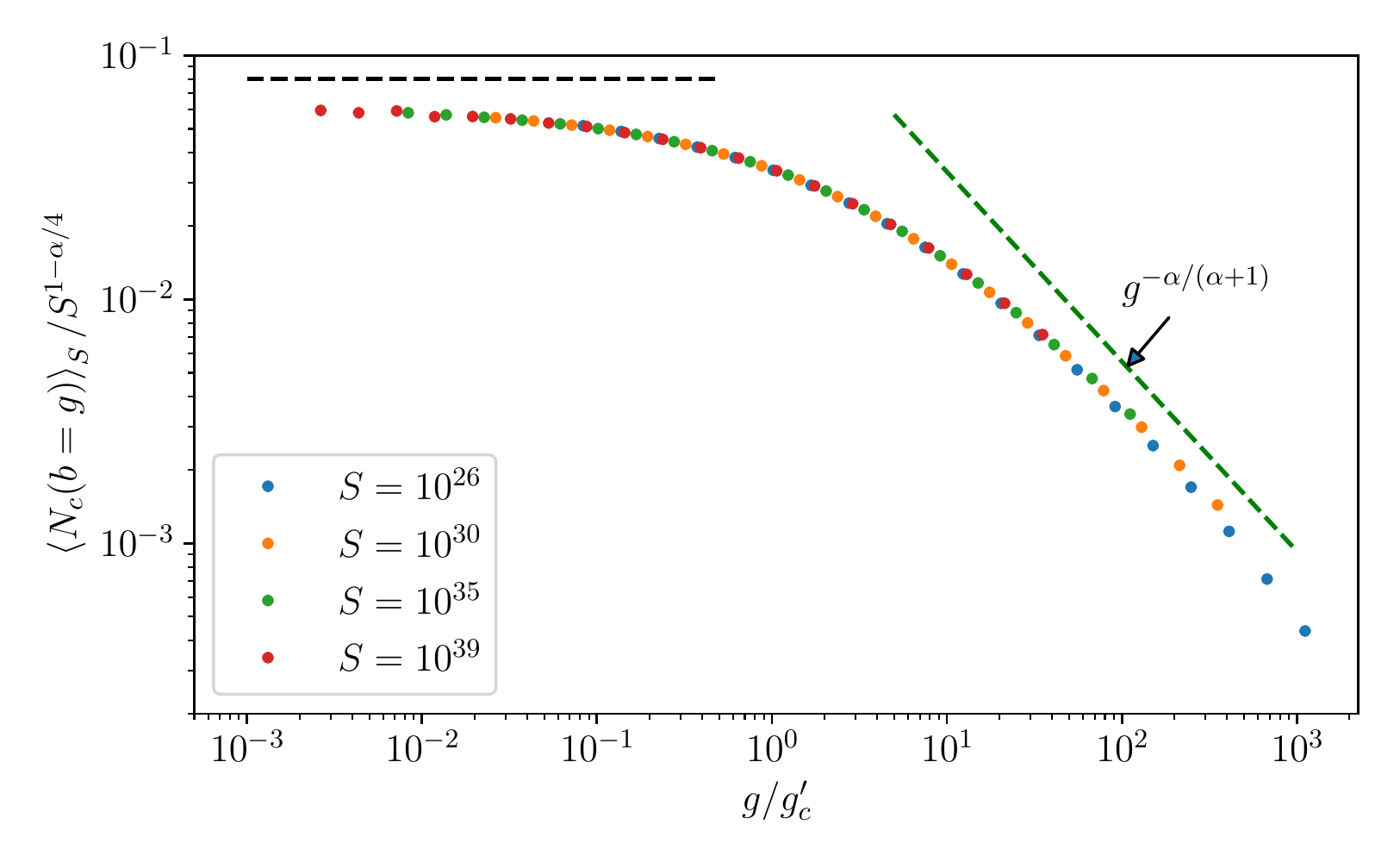}
    \caption{Cluster number (gap distribution) with $\alpha = 3.5, d= 1$. We solved the instanton equation for $b/\epsilon\in [60, 2\times10^6]$, and $m^2 = 10^{-13}, 10^{-15},10^{-17}, 10^{-19}$ as well as $1.5 m^2$, in order to extract $S$-conditioned averages at $S = m^{-4}$, using \eqref{eq:OSm2}. The data collapse confirms the prediction \eqref{eq:Ncalphalarge} as well as that of the gap distribution for $g  > g_c$.  }
    \label{fig:SR}
\end{figure}

\item $\alpha > 4$. The short-range solution decays more slowly than the near-plateau approximation $|x|^{-\alpha/2}$. This implies that the plateau contribution cannot dominate at large distances for any value of $b$. The solution is always the short range scale invariant one up to the cutoff $\xi_{\text{SR}}$, with no significant dependence on $b$ at long distances:
\begin{empheq}[box=\widefbox]{equation}
    F = F_{\text{SR}} \sim |x|^{-2} \,,\, |x| \ll \xi_{\text{SR}}
\end{empheq}
Therefore we have a short-range behavior for the extension and area
\begin{equation}
  \left< \ell(b) \right>_S \,,\,  \left< \mathcal{A}(b) \right>_S  \sim  \partial_{m^2} \int F(x) \mathrm{d}^d x \sim \xi_{\text{SR}}^d  \,. \label{eq:SRellA}
\end{equation}
Because there is no longer a $b$ dependence, there are no longer a large number of clusters. In other words, the short-range physics completely takes over when $\alpha > 4$, confirming the simple argument of the main text.
\end{enumerate}

\subsection{Relation and application to avalanches}

In this section we discuss the above results from the perspective of depinning avalanches with long-range elasticity. We assume $\alpha < 2$ unless otherwise stated. 

We recall that an avalanche of an elastic interface is described by the following equation of motion governing the interface position $u(x,t)$:
\begin{equation}
    \partial_t u = (\mathcal{D}^\alpha - m^2 )u + f(x, u) + \delta(t-t_0)\delta(x-x_0) m^2 w \,.
\end{equation}
Here $ \mathcal{D}^\alpha$ is the elastic interaction (plus a mass term), $m^2$ is a mass (it introduces a cut off to the avalanche size, in a similar way as does $1-R_0$), $f$ is the {quenched random force from the disordered medium, which for realistic models is usually short range in space. The last} term is a localized kick in order to trigger an avalanche. The mean-field approximation, {which can be justified in dimension $d>2 \alpha$}, consists in replacing $f$ by an independent Brownian motion in $u$ for each $x$, such that:
\begin{equation}
 \left< (f(x,u') - f(x,u))^2 \right> = |u-u'| \,.
\end{equation}
Then we obtain the Brownian force model, which is exactly solvable~\cite{pldw12,pldw13}. If we place the interface at an equilibrium before the kick, the interface will move forward and eventually stop at a further equilibrium. The displacement is the avalanche. Its total size $S = \int \mathrm{d}^d x \mathrm{d} t \partial_t u$ is defined as the integrated displacement. 

It is known that the Brownian force model is equivalent to a continuum limit of the epidemic model discussed in this work~\cite{pierrenew}. We will comment on this below. Now, let us take this mapping for granted and discuss the results of this work from the perspective of mean-field avalanches. 

\subsubsection{Applications}
\textbf{Roughness.} In the mapping to the Brownian force model the total number of infections $S$ corresponds to the total size of the avalanche. Its distribution is fixed by the BGW law: $P(S) \sim S^{-\tau}$, with $\tau = 3/2$. Now, the Narayan-Fisher~\cite{narayan,zapperi98,toussaint06} scaling relation 
(see also extensions in \cite{DobrinevskiLeDoussalWiese2014a}) 
relates $\tau$ to the roughness exponent $\zeta$ ($u\sim x^\zeta$):
\begin{equation}
    \tau = 2 - \frac{\alpha}{d+\zeta} \,,\, \alpha > d/2 \,.
\end{equation}
Plugging in $\tau = 3/2$, we find $\zeta = 2 \alpha - d$. When $\alpha < d/2$, the above relation does not apply, and $\zeta = 0$ instead. How does this relate to our results?
\begin{itemize}
    \item When $\alpha < d/2$, $\zeta = 0$ corresponds to our finding that $\left< \ell \right>_S ,\left< \mathcal{A} \right>_S  \sim S $: The avalanche is flat instead of rough, so that its extension/area is proportional to its size. 
    \item When $\alpha > d/2$, the roughness exponent relates the bulk extent with the size in a standard way:
    \begin{equation}
        S \sim \xi^{2\alpha} \sim \xi^{d + \zeta}  \,. \label{eq:self-affine}
    \end{equation}
    This relation is usually interpreted as attesting the self-affinity of the rough avalanche.  However, the scaling relation between the extension/area and $S$ is given by \eqref{eq:self-affine} (as one expects) only if $d/2<\alpha<d$. In 1D, and when $\alpha \in (1,2)$, the mean extension $\left< \ell \right> \sim S^{1/(1+\alpha)}$ is much larger than the bulk extent $S^{1/(2\alpha)}$, and is dominated by many small clusters in the outskirt ($\left< N_c\right> \sim S^{1/(1+\alpha)}$ as well). These clusters are not expected to be self-affine. Yet, we expect the largest clusters in the bulk to have size $\sim \xi$ and be self-affine with roughness $\zeta$. Probing these clusters directly is beyond the reach of the present approach.
\end{itemize}

\textbf{A few more exponents.} 
The nontrivial cluster number exponent $\chi$ gives rise to a few other predictions (conjectures), for $\alpha \in (d/2,d)$.
\begin{itemize}
    \item By assuming self-affinity of the bulk clusters, we can relate the clusters' size $S_c$ with their extension/area via $S_c \sim \ell_c^{2\alpha}$, $S_c \sim \mathcal{A}^\alpha$. Combining this with the conjecture on the distribution of $\ell_c$ and $\mathcal{A}_c$ (given in the main text), we can conjecture the cluster size distribution:
    \begin{equation}\label{eq:PSc}
        P(S_c) \sim S_c^{-\frac{\chi}{2\alpha}-1} \,.
    \end{equation}
    \item Using the scaling relation $N_c \sim \xi^{\chi} \sim S^{\frac{\chi}{2\alpha}}$ and the BGW law for $S$, we obtain a conjecture on the distribution of cluster number:
    \begin{equation}\label{eq:PNc}
        P(N_c) \sim N_c^{-\frac{\alpha}{\chi} - 1} \,.
    \end{equation}
    The exponent goes to $3/2$ as $\alpha \to d/2$ (since $\chi \to d$). Away from that limit, the exponent is larger than $3/2$. Of course this is not in contradiction with the numerical observation of Ref.~\cite{lepriol21} that $P(N_c) \sim N_c^{-3/2}$ for all $\alpha$ in \textit{non-mean-field} avalanches. The robustness of the exponent $3/2$ is probably a result of loop corrections. 
\end{itemize}

\subsubsection{Remarks on the mapping and the continuum limit}
 We now briefly discuss the mapping between the epidemic model and the Brownian force model, focusing on the continuum limit involved~\cite{pierrenew}. From the point of view of the epidemic model, the first step of this continuum limit is to take the infection and recovery rate to be large:
\begin{equation}
    \beta = N \gg 1 \,,\, \gamma = N  + m^2  \,.
\end{equation}
Thereby, the instanton equation $\partial_t F = \mathcal{D}^\alpha F - N  F^2 - m^2 F $ acquires a large parameter. The second step is to remove the large parameter from the instanton equation, by changing the observable. Recall that in our approach, $E = 1-F$ is the probability that the $b$-neighborhood of the origin is \textit{not} visited by an epidemic starting from $x$~\eqref{eq:E}. We can soften this ``hardcore repulsion''  observable as follows: 
\begin{equation}
    G(x,t) := \left<  \exp\left(-\int_{-t}^0 \mathrm{d} s \sum_{i=1}^{I(s)}  \frac1N \lambda(x_i(s), s)  \right)  \right> \,,
\end{equation}
where $\lambda(x,t)$ is a function of space, and the epidemic starts at $-t$ with a single infected individual at $x$ (The choice of time coordinate looks awkward but necessary for a backward recursion to work.). In particular, taking $\lambda(x,t) = C N \theta(b / 2 - \Vert x \Vert)$, with $C\to \infty$, we recover $E$~\eqref{eq:E}. It is not hard to see that $G$ satisfies a similar backward recursion as $E$:
\begin{equation}
    \partial_t G = \mathcal{D}^\alpha G +  N (1-G)^2 +  m^2 (1-G) - \frac1N \lambda G \,.
\end{equation}
Finally, setting $G = \exp(- \tilde{u}/N)$, and keeping the leading order in $N$, we obtain 
\begin{equation}
    \partial_t \tilde{u} = \mathcal{D}^\alpha \tilde{u} - \tilde{u}^2 - {m}^2 \tilde{u} + \lambda \,. \label{eq:BFM-instanton}
\end{equation}
This is the instanton equation in the Brownian force model~\cite{pldw13} (often a different sign convention is used where $\tilde{u} \to -\tilde{u}$
and $\lambda \to - \lambda$). Note that it is very similar to our instanton equation for $F$. However the differences are important, in particular the physical interpretation is distinct. In the Brownian force model, the velocity of the interface corresponds to the (scaled) density of infected individuals
\begin{equation}
    {\partial_t u}(x,t)= \sum_{i=1}^{I(t)} \frac1N \delta(x_i(t) - x) \,.
\end{equation}
In particular, if we set $\lambda(x,t) \equiv \lambda$,
\begin{equation}
    e^{- \tilde{u} m^2 w } = G^{I(0)} = \left<\exp\left(- \lambda \int  \, {\partial_t u} \mathrm{d}^d x \mathrm{d} t\right)  \right> \,,\, I(0) = m^2 w N
\end{equation}
is the generating function of the total avalanche size, usually called $S$ in the avalanche literature. It corresponds to th e (scaled) total lifetime of all infected individuals (until recovery), if we start with $I(0)$ infected individuals. This is in turn proportional to the total number of infections. So it is reasonable to call the latter $S$ as well, as we did in the main text.  

We remark that, once the continuum limit is taken, it becomes not obvious to speak about clusters. Indeed, putting a source term in \eqref{eq:BFM-instanton}, for example $\lambda = \lambda_0 \theta(b/2-\Vert x \Vert)$, amounts to imposing a {finite} penalty/fugacity for the time spent by infected individuals (or avalanche activities) in the $b$-neighborhood. To define the emptyness probability one needs to send $\lambda_0 \to \infty$. This has been done in the case of short-range elasticity to calculate, for example, the extension of the avalanche~\cite{pldw13,Thiery_2015,pierrenewBFM}. But such a limit would be ill-behaved with long-range elasticity, since $\tilde{u}$ would diverge everywhere. By contrast in our model, the instanton equation for $F$ does not have a source term but rather a ``boundary condition'', $F\vert_{\Vert x \Vert< b/2} = 1$. Similar boundary conditions for $\tilde{u}$~\eqref{eq:BFM-instanton} also appeared in the continuum limit, but the interpretation is different, see~\cite{pierrenew} Appendix D2, and references therein.

{{\it Remark: reduction in dimension}. The 1D model can be obtained from the 2D one by a projection.
Writing $(x_i(t),y_i(t))$ the coordinates of the infected individuals in the 2D model, then the process $x_i(t)$ is the 1D model with the same infection/recovery rates and the jump rate $p_\alpha(x-x') dx' dt$ where $p_\alpha(x-x')= \int dy'/((x-x')^2+(y-y')^2)^{\frac{2+\alpha}{2}} \sim |x-x'|^{-(1+\alpha)}$, up to an irrelevant change in the precise cutoff function at scale $\epsilon$. A similar reduction in dimension holds for the Brownian force model, see e.g. \cite{pierrenewBFM} Sec. IA. This reduction implies a number of bounds between 2D and 1D quantities with the same $\alpha$. For example, the number of clusters increase with the dimension: $\left< N_c(b) \right>_{\alpha, \text{2D}} \ge \left< N_c(b) \right>_{\alpha, \text{1D}}$. We can check that our asymptotic results satisfy this bound. This bound is not tight: for instance when $\alpha \in (1/2,1)$, $\left<N_c(b=1) \right>\sim S$ in 2D but $\left<N_c(b=1) \right>\sim S^{\chi/(2\alpha)} \ll S$ in 1D. 
}

\subsection{Numerical methods}
In this section we describe the numerical techniques used in the extensive solution of the instanton equation. We restrict to $ d = 1$.

\subsubsection{Logarithmic discretisation}
To solve the instanton equation numerically, we need to discretise space. The standard uniform mesh is not sufficient to attain the large length scale where the asymptotic behaviors predicted above can be clearly observed. We take a logarithmic mesh
\begin{equation}
    x_n = \epsilon \exp(n \delta) \,,\, n = 0, 1, 2, \dots, N.
\end{equation}
Usually, we choose $\epsilon = 1$, the mesh size $\delta = 0.1$ or $\delta = 0.2$ is sufficient fine, and $N = 200\sim 400$ should be chosen appropriately to avoid finite size effects. For example, for the critical regime, one should see a clear mass cutoff. The value of $b$ is chosen to be $2x_n$ for some $n$. 

Then, we approximate $F$ to be piece-wise linear in the intervals $|x| \in [-x_0, x_0]$ and $|x| \in [x_n, x_{n+1}]$. We also assume $F(x) = F(-x)$. When calculating $(\mathcal{D}^\alpha F)(x)$ in the interval $|x| \in (x_n, x_{n+1})$, we approximate $x$ to be the middle point $z_n = (x_n + x_{n+1}) / 2$ for the whole interval:
\begin{align}
   & (\mathcal{D}^\alpha F)(x \in [x_n, x_{n+1}]) = 
    \int p_\alpha(z_n - y) (F(y)- F(x)) \mathrm{d} y = \sum_{m=-1}^N (F(z_m) - F(z_n)) K_{nm} \,,\, \\
&    K_{nm}  = \int_{|y| \in [x_m, x_{m+1}]} p_\alpha(z_n - y) \mathrm{d} y 
\end{align}
where we have set $z_{-1} = 0$ (it corresponds to the interval $[-x_0, x_0]$. The matrix elements can be explicitly calculated using $p_\alpha(x) = |x|^{-1-\alpha}$, for $n\ne m$. We do not need the ones for $n = m$. 

It should be noted that the logarithmic mesh amounts to different way of cutting off $p_\alpha(x)$ at short distances: the cutoff effectively depends on the position. This has an undesired effect for $\alpha > 2$: we do not get the short distance term $\propto |k|^2$ automatically, and have to add it by hand: $K \to K + K_{\text{SR}}$, where $ K_{\text{SR}}$ is the above matrix with $\alpha = 2$, from which we remove all the elements with distance $> 1$ away from the diagonal (so that $K_{\text{SR}} $ is tri-diagonal). 

\subsubsection{Iteration scheme for stationary solution}
In critical regime, we need to find the stationary ($t\to\infty$) solution to the instanton equation. We find this by iteration. For this, we can write the instanton equation as 
\begin{equation}
    \sum_m K_{nm} F(z_m) = (m^2 +  \sum_m K_{nm} ) F(z_n) + F(z_n)^2 \,, |z_n| > b/2 \,.
\end{equation}
Then the iteration scheme is as follows:
\begin{align}
&F_0(z_n) = \theta(b/2 - z_n) \,,\, \\
&(m^2 +  \sum_m K_{nm} ) F_{j+1}(z_n) + F_{j+1}(z_n)^2 =   \sum_m K_{nm} F_j(z_m)  \,,\, z_n > b/2 \,.
\end{align}
that is, for each iteration, we compute the matrix multiplication of the RHS, and then solve the quadratic equation for $F_{j+1}$ (we pick the positive solution). It is not hard to show that $F_j(z)$ increases with $j$. Since $F\le 1$ is also bounded from above, the (point-wise) convergence of this procedure is guaranteed. In practice, a few hundred iterations provide a sufficient convergence for all the tests we presented. This corresponds to no more than to a couple of minutes of calculation on a consumer laptop in order to generate each plot of this paper from scratch.

In the supercritical regime, we solve the time-dependent instanton equation by the Euler scheme with $\delta t = 0.01$.

\subsubsection{Code availability}
The code used to generate all the plots of this paper is available by following this link: \href{https://github.com/xcao-phys/cluster}{https://github.com/xcao-phys/cluster}.

\end{widetext}

\bibliography{ref.bib}

%merlin.mbs apsrev4-1.bst 2010-07-25 4.21a (PWD, AO, DPC) hacked
%Control: key (0)
%Control: author (0) dotless jnrlst
%Control: editor formatted (1) identically to author
%Control: production of article title (0) allowed
%Control: page (1) range
%Control: year (0) verbatim
%Control: production of eprint (0) enabled
\begin{thebibliography}{62}%
\makeatletter
\providecommand \@ifxundefined [1]{%
 \@ifx{#1\undefined}
}%
\providecommand \@ifnum [1]{%
 \ifnum #1\expandafter \@firstoftwo
 \else \expandafter \@secondoftwo
 \fi
}%
\providecommand \@ifx [1]{%
 \ifx #1\expandafter \@firstoftwo
 \else \expandafter \@secondoftwo
 \fi
}%
\providecommand \natexlab [1]{#1}%
\providecommand \enquote  [1]{``#1''}%
\providecommand \bibnamefont  [1]{#1}%
\providecommand \bibfnamefont [1]{#1}%
\providecommand \citenamefont [1]{#1}%
\providecommand \href@noop [0]{\@secondoftwo}%
\providecommand \href [0]{\begingroup \@sanitize@url \@href}%
\providecommand \@href[1]{\@@startlink{#1}\@@href}%
\providecommand \@@href[1]{\endgroup#1\@@endlink}%
\providecommand \@sanitize@url [0]{\catcode `\\12\catcode `\$12\catcode
  `\&12\catcode `\#12\catcode `\^12\catcode `\_12\catcode `\%12\relax}%
\providecommand \@@startlink[1]{}%
\providecommand \@@endlink[0]{}%
\providecommand \url  [0]{\begingroup\@sanitize@url \@url }%
\providecommand \@url [1]{\endgroup\@href {#1}{\urlprefix }}%
\providecommand \urlprefix  [0]{URL }%
\providecommand \Eprint [0]{\href }%
\providecommand \doibase [0]{http://dx.doi.org/}%
\providecommand \selectlanguage [0]{\@gobble}%
\providecommand \bibinfo  [0]{\@secondoftwo}%
\providecommand \bibfield  [0]{\@secondoftwo}%
\providecommand \translation [1]{[#1]}%
\providecommand \BibitemOpen [0]{}%
\providecommand \bibitemStop [0]{}%
\providecommand \bibitemNoStop [0]{.\EOS\space}%
\providecommand \EOS [0]{\spacefactor3000\relax}%
\providecommand \BibitemShut  [1]{\csname bibitem#1\endcsname}%
\let\auto@bib@innerbib\@empty
%</preamble>
\bibitem [{Bie()}]{Bienayme}%
  \BibitemOpen
  \href@noop {} {}\bibinfo {note} {I.J. Bienaym\'e, {\it De la loi de
  multiplication et de la dur\'ee des familles}, Soc. Philomat. Paris Extraits,
  S\'er 5, 37-39 (1845). L'institut 589, 13:131-132. Reprinted in D.G. Kendall,
  {\it The genealogy of genealogy: Branching processes before (and after)
  1873}, Bull. London. Math. Soc. 7:225-253 (1975).}\BibitemShut {Stop}%
\bibitem [{WG()}]{WG}%
  \BibitemOpen
  \href@noop {} {}\bibinfo {note} {H. Watson and F. Galton, {\it On the
  probability of the extinction of families}, J. Anthropol. Inst. G. B. Irel.
  4, 138 (1875).}\BibitemShut {Stop}%
\bibitem [{\citenamefont {Alessandro}\ \emph {et~al.}(1990)\citenamefont
  {Alessandro}, \citenamefont {Beatrice}, \citenamefont {Bertotti},\ and\
  \citenamefont {Montorsi}}]{abbm}%
  \BibitemOpen
  \bibfield  {author} {\bibinfo {author} {\bibfnamefont {Bruno}\ \bibnamefont
  {Alessandro}}, \bibinfo {author} {\bibfnamefont {Cinzia}\ \bibnamefont
  {Beatrice}}, \bibinfo {author} {\bibfnamefont {Giorgio}\ \bibnamefont
  {Bertotti}}, \ and\ \bibinfo {author} {\bibfnamefont {Arianna}\ \bibnamefont
  {Montorsi}},\ }\bibfield  {title} {\enquote {\bibinfo {title} {Domain‐wall
  dynamics and barkhausen effect in metallic ferromagnetic materials. i.
  theory},}\ }\href {\doibase 10.1063/1.346423} {\bibfield  {journal} {\bibinfo
   {journal} {Journal of Applied Physics}\ }\textbf {\bibinfo {volume} {68}},\
  \bibinfo {pages} {2901--2907} (\bibinfo {year} {1990})}\BibitemShut {NoStop}%
\bibitem [{\citenamefont {Bramson}(1978)}]{bramson}%
  \BibitemOpen
  \bibfield  {author} {\bibinfo {author} {\bibfnamefont {Maury~D.}\
  \bibnamefont {Bramson}},\ }\bibfield  {title} {\enquote {\bibinfo {title}
  {Maximal displacement of branching brownian motion},}\ }\href {\doibase
  https://doi.org/10.1002/cpa.3160310502} {\bibfield  {journal} {\bibinfo
  {journal} {Communications on Pure and Applied Mathematics}\ }\textbf
  {\bibinfo {volume} {31}},\ \bibinfo {pages} {531--581} (\bibinfo {year}
  {1978})}\BibitemShut {NoStop}%
\bibitem [{\citenamefont {Slade}(2002)}]{slade2002scaling}%
  \BibitemOpen
  \bibfield  {author} {\bibinfo {author} {\bibfnamefont {Gordon}\ \bibnamefont
  {Slade}},\ }\bibfield  {title} {\enquote {\bibinfo {title} {Scaling limits
  and super-brownian motion},}\ }\href
  {https://www.ams.org/journals/notices/200209/fea-sladecolor.pdf?trk=200209fea-sladecolor&cat=collection}
  {\bibfield  {journal} {\bibinfo  {journal} {Notices AMS}\ }\textbf {\bibinfo
  {volume} {49}},\ \bibinfo {pages} {1056--1067} (\bibinfo {year}
  {2002})}\BibitemShut {NoStop}%
\bibitem [{\citenamefont {Brunet}\ and\ \citenamefont
  {Derrida}(2009)}]{Brunet_2009}%
  \BibitemOpen
  \bibfield  {author} {\bibinfo {author} {\bibfnamefont {{\'{E}}.}~\bibnamefont
  {Brunet}}\ and\ \bibinfo {author} {\bibfnamefont {B.}~\bibnamefont
  {Derrida}},\ }\bibfield  {title} {\enquote {\bibinfo {title} {Statistics at
  the tip of a branching random walk and the delay of traveling waves},}\
  }\href {\doibase 10.1209/0295-5075/87/60010} {\bibfield  {journal} {\bibinfo
  {journal} {{EPL} (Europhysics Letters)}\ }\textbf {\bibinfo {volume} {87}},\
  \bibinfo {pages} {60010} (\bibinfo {year} {2009})}\BibitemShut {NoStop}%
\bibitem [{\citenamefont {Arguin}\ \emph {et~al.}(2013)\citenamefont {Arguin},
  \citenamefont {Bovier},\ and\ \citenamefont {Kistler}}]{arguin}%
  \BibitemOpen
  \bibfield  {author} {\bibinfo {author} {\bibfnamefont {Louis-Pierre}\
  \bibnamefont {Arguin}}, \bibinfo {author} {\bibfnamefont {Anton}\
  \bibnamefont {Bovier}}, \ and\ \bibinfo {author} {\bibfnamefont {Nicola}\
  \bibnamefont {Kistler}},\ }\bibfield  {title} {\enquote {\bibinfo {title}
  {The extremal process of branching brownian motion},}\ }\href {\doibase
  10.1007/s00440-012-0464-x} {\bibfield  {journal} {\bibinfo  {journal}
  {Probability Theory and Related Fields}\ }\textbf {\bibinfo {volume} {157}},\
  \bibinfo {pages} {535--574} (\bibinfo {year} {2013})}\BibitemShut {NoStop}%
\bibitem [{\citenamefont {Dumonteil}\ \emph {et~al.}(2013)\citenamefont
  {Dumonteil}, \citenamefont {Majumdar}, \citenamefont {Rosso},\ and\
  \citenamefont {Zoia}}]{dumonteil13extent}%
  \BibitemOpen
  \bibfield  {author} {\bibinfo {author} {\bibfnamefont {Eric}\ \bibnamefont
  {Dumonteil}}, \bibinfo {author} {\bibfnamefont {Satya~N.}\ \bibnamefont
  {Majumdar}}, \bibinfo {author} {\bibfnamefont {Alberto}\ \bibnamefont
  {Rosso}}, \ and\ \bibinfo {author} {\bibfnamefont {Andrea}\ \bibnamefont
  {Zoia}},\ }\bibfield  {title} {\enquote {\bibinfo {title} {Spatial extent of
  an outbreak in animal epidemics},}\ }\href {\doibase 10.1073/pnas.1213237110}
  {\bibfield  {journal} {\bibinfo  {journal} {Proceedings of the National
  Academy of Sciences}\ }\textbf {\bibinfo {volume} {110}},\ \bibinfo {pages}
  {4239--4244} (\bibinfo {year} {2013})}\BibitemShut {NoStop}%
\bibitem [{\citenamefont {Ramola}\ \emph {et~al.}(2015)\citenamefont {Ramola},
  \citenamefont {Majumdar},\ and\ \citenamefont {Schehr}}]{ramola}%
  \BibitemOpen
  \bibfield  {author} {\bibinfo {author} {\bibfnamefont {Kabir}\ \bibnamefont
  {Ramola}}, \bibinfo {author} {\bibfnamefont {Satya~N.}\ \bibnamefont
  {Majumdar}}, \ and\ \bibinfo {author} {\bibfnamefont {Gr\'egory}\
  \bibnamefont {Schehr}},\ }\bibfield  {title} {\enquote {\bibinfo {title}
  {Spatial extent of branching brownian motion},}\ }\href {\doibase
  10.1103/PhysRevE.91.042131} {\bibfield  {journal} {\bibinfo  {journal} {Phys.
  Rev. E}\ }\textbf {\bibinfo {volume} {91}},\ \bibinfo {pages} {042131}
  (\bibinfo {year} {2015})}\BibitemShut {NoStop}%
\bibitem [{\citenamefont {Ramola}\ \emph {et~al.}(2014)\citenamefont {Ramola},
  \citenamefont {Majumdar},\ and\ \citenamefont {Schehr}}]{ramola2}%
  \BibitemOpen
  \bibfield  {author} {\bibinfo {author} {\bibfnamefont {Kabir}\ \bibnamefont
  {Ramola}}, \bibinfo {author} {\bibfnamefont {Satya~N.}\ \bibnamefont
  {Majumdar}}, \ and\ \bibinfo {author} {\bibfnamefont {Gr\'egory}\
  \bibnamefont {Schehr}},\ }\bibfield  {title} {\enquote {\bibinfo {title}
  {Universal order and gap statistics of critical branching brownian motion},}\
  }\href {\doibase 10.1103/PhysRevLett.112.210602} {\bibfield  {journal}
  {\bibinfo  {journal} {Phys. Rev. Lett.}\ }\textbf {\bibinfo {volume} {112}},\
  \bibinfo {pages} {210602} (\bibinfo {year} {2014})}\BibitemShut {NoStop}%
\bibitem [{\citenamefont {Suarez}\ \emph {et~al.}(2001)\citenamefont {Suarez},
  \citenamefont {Holway},\ and\ \citenamefont {Case}}]{01ants}%
  \BibitemOpen
  \bibfield  {author} {\bibinfo {author} {\bibfnamefont {Andrew~V.}\
  \bibnamefont {Suarez}}, \bibinfo {author} {\bibfnamefont {David~A.}\
  \bibnamefont {Holway}}, \ and\ \bibinfo {author} {\bibfnamefont {Ted~J.}\
  \bibnamefont {Case}},\ }\bibfield  {title} {\enquote {\bibinfo {title}
  {Patterns of spread in biological invasions dominated by long-distance jump
  dispersal: Insights from argentine ants},}\ }\href {\doibase
  10.1073/pnas.98.3.1095} {\bibfield  {journal} {\bibinfo  {journal}
  {Proceedings of the National Academy of Sciences}\ }\textbf {\bibinfo
  {volume} {98}},\ \bibinfo {pages} {1095--1100} (\bibinfo {year}
  {2001})}\BibitemShut {NoStop}%
\bibitem [{\citenamefont {Brown}\ and\ \citenamefont
  {Hovmøller}(2002)}]{02aerial}%
  \BibitemOpen
  \bibfield  {author} {\bibinfo {author} {\bibfnamefont {James K.~M.}\
  \bibnamefont {Brown}}\ and\ \bibinfo {author} {\bibfnamefont {Mogens~S.}\
  \bibnamefont {Hovmøller}},\ }\bibfield  {title} {\enquote {\bibinfo {title}
  {Aerial dispersal of pathogens on the global and continental scales and its
  impact on plant disease},}\ }\href {\doibase 10.1126/science.1072678}
  {\bibfield  {journal} {\bibinfo  {journal} {Science}\ }\textbf {\bibinfo
  {volume} {297}},\ \bibinfo {pages} {537--541} (\bibinfo {year}
  {2002})}\BibitemShut {NoStop}%
\bibitem [{\citenamefont {Nathan}(2006)}]{nathanplant}%
  \BibitemOpen
  \bibfield  {author} {\bibinfo {author} {\bibfnamefont {Ran}\ \bibnamefont
  {Nathan}},\ }\bibfield  {title} {\enquote {\bibinfo {title} {Long-distance
  dispersal of plants},}\ }\href {\doibase 10.1126/science.1124975} {\bibfield
  {journal} {\bibinfo  {journal} {Science}\ }\textbf {\bibinfo {volume}
  {313}},\ \bibinfo {pages} {786--788} (\bibinfo {year} {2006})}\BibitemShut
  {NoStop}%
\bibitem [{\citenamefont {Brockmann}(2009)}]{brockmann}%
  \BibitemOpen
  \bibfield  {author} {\bibinfo {author} {\bibfnamefont {Dirk}\ \bibnamefont
  {Brockmann}},\ }\enquote {\bibinfo {title} {Human mobility and spatial
  disease dynamics},}\ in\ \href {\doibase
  https://doi.org/10.1002/9783527628001.ch1} {\emph {\bibinfo {booktitle}
  {Reviews of Nonlinear Dynamics and Complexity}}}\ (\bibinfo  {publisher}
  {John Wiley \& Sons, Ltd},\ \bibinfo {year} {2009})\ Chap.~\bibinfo {chapter}
  {1}, pp.\ \bibinfo {pages} {1--24}\BibitemShut {NoStop}%
\bibitem [{\citenamefont {Gonz{\'a}lez}\ \emph {et~al.}(2008)\citenamefont
  {Gonz{\'a}lez}, \citenamefont {Hidalgo},\ and\ \citenamefont
  {Barab{\'a}si}}]{gonza08nature}%
  \BibitemOpen
  \bibfield  {author} {\bibinfo {author} {\bibfnamefont {Marta~C.}\
  \bibnamefont {Gonz{\'a}lez}}, \bibinfo {author} {\bibfnamefont
  {C{\'e}sar~A.}\ \bibnamefont {Hidalgo}}, \ and\ \bibinfo {author}
  {\bibfnamefont {Albert-L{\'a}szl{\'o}}\ \bibnamefont {Barab{\'a}si}},\
  }\bibfield  {title} {\enquote {\bibinfo {title} {Understanding individual
  human mobility patterns},}\ }\href {\doibase 10.1038/nature06958} {\bibfield
  {journal} {\bibinfo  {journal} {Nature}\ }\textbf {\bibinfo {volume} {453}},\
  \bibinfo {pages} {779--782} (\bibinfo {year} {2008})}\BibitemShut {NoStop}%
\bibitem [{\citenamefont {Perlekar}\ \emph {et~al.}(2010)\citenamefont
  {Perlekar}, \citenamefont {Benzi}, \citenamefont {Nelson},\ and\
  \citenamefont {Toschi}}]{perlekar10turbulent}%
  \BibitemOpen
  \bibfield  {author} {\bibinfo {author} {\bibfnamefont {Prasad}\ \bibnamefont
  {Perlekar}}, \bibinfo {author} {\bibfnamefont {Roberto}\ \bibnamefont
  {Benzi}}, \bibinfo {author} {\bibfnamefont {David~R.}\ \bibnamefont
  {Nelson}}, \ and\ \bibinfo {author} {\bibfnamefont {Federico}\ \bibnamefont
  {Toschi}},\ }\bibfield  {title} {\enquote {\bibinfo {title} {Population
  dynamics at high reynolds number},}\ }\href {\doibase
  10.1103/PhysRevLett.105.144501} {\bibfield  {journal} {\bibinfo  {journal}
  {Phys. Rev. Lett.}\ }\textbf {\bibinfo {volume} {105}},\ \bibinfo {pages}
  {144501} (\bibinfo {year} {2010})}\BibitemShut {NoStop}%
\bibitem [{\citenamefont {Colizza}\ \emph {et~al.}(2006)\citenamefont
  {Colizza}, \citenamefont {Barrat}, \citenamefont {Barthélemy},\ and\
  \citenamefont {Vespignani}}]{barrat}%
  \BibitemOpen
  \bibfield  {author} {\bibinfo {author} {\bibfnamefont {Vittoria}\
  \bibnamefont {Colizza}}, \bibinfo {author} {\bibfnamefont {Alain}\
  \bibnamefont {Barrat}}, \bibinfo {author} {\bibfnamefont {Marc}\ \bibnamefont
  {Barthélemy}}, \ and\ \bibinfo {author} {\bibfnamefont {Alessandro}\
  \bibnamefont {Vespignani}},\ }\bibfield  {title} {\enquote {\bibinfo {title}
  {The role of the airline transportation network in the prediction and
  predictability of global epidemics},}\ }\href {\doibase
  10.1073/pnas.0510525103} {\bibfield  {journal} {\bibinfo  {journal}
  {Proceedings of the National Academy of Sciences}\ }\textbf {\bibinfo
  {volume} {103}},\ \bibinfo {pages} {2015--2020} (\bibinfo {year} {2006})},\
  \Eprint
  {http://arxiv.org/abs/https://www.pnas.org/doi/pdf/10.1073/pnas.0510525103}
  {https://www.pnas.org/doi/pdf/10.1073/pnas.0510525103} \BibitemShut {NoStop}%
\bibitem [{\citenamefont {Rice}(1985)}]{rice85}%
  \BibitemOpen
  \bibfield  {author} {\bibinfo {author} {\bibfnamefont {J.~R.}\ \bibnamefont
  {Rice}},\ }\bibfield  {title} {\enquote {\bibinfo {title} {{First-Order
  Variation in Elastic Fields Due to Variation in Location of a Planar Crack
  Front}},}\ }\href {\doibase 10.1115/1.3169103} {\bibfield  {journal}
  {\bibinfo  {journal} {Journal of Applied Mechanics}\ }\textbf {\bibinfo
  {volume} {52}},\ \bibinfo {pages} {571--579} (\bibinfo {year}
  {1985})}\BibitemShut {NoStop}%
\bibitem [{\citenamefont {Gao}\ and\ \citenamefont {Rice}(1989)}]{gao89crack}%
  \BibitemOpen
  \bibfield  {author} {\bibinfo {author} {\bibfnamefont {Huajian}\ \bibnamefont
  {Gao}}\ and\ \bibinfo {author} {\bibfnamefont {James~R.}\ \bibnamefont
  {Rice}},\ }\bibfield  {title} {\enquote {\bibinfo {title} {{A First-Order
  Perturbation Analysis of Crack Trapping by Arrays of Obstacles}},}\ }\href
  {\doibase 10.1115/1.3176178} {\bibfield  {journal} {\bibinfo  {journal}
  {Journal of Applied Mechanics}\ }\textbf {\bibinfo {volume} {56}},\ \bibinfo
  {pages} {828--836} (\bibinfo {year} {1989})}\BibitemShut {NoStop}%
\bibitem [{\citenamefont {Tanguy}\ \emph {et~al.}(1998)\citenamefont {Tanguy},
  \citenamefont {Gounelle},\ and\ \citenamefont {Roux}}]{tanguy98crack}%
  \BibitemOpen
  \bibfield  {author} {\bibinfo {author} {\bibfnamefont {Anne}\ \bibnamefont
  {Tanguy}}, \bibinfo {author} {\bibfnamefont {Matthieu}\ \bibnamefont
  {Gounelle}}, \ and\ \bibinfo {author} {\bibfnamefont {St\'ephane}\
  \bibnamefont {Roux}},\ }\bibfield  {title} {\enquote {\bibinfo {title} {From
  individual to collective pinning: Effect of long-range elastic
  interactions},}\ }\href {\doibase 10.1103/PhysRevE.58.1577} {\bibfield
  {journal} {\bibinfo  {journal} {Phys. Rev. E}\ }\textbf {\bibinfo {volume}
  {58}},\ \bibinfo {pages} {1577--1590} (\bibinfo {year} {1998})}\BibitemShut
  {NoStop}%
\bibitem [{\citenamefont {Bonamy}\ \emph {et~al.}(2008)\citenamefont {Bonamy},
  \citenamefont {Santucci},\ and\ \citenamefont {Ponson}}]{bonamy08crack}%
  \BibitemOpen
  \bibfield  {author} {\bibinfo {author} {\bibfnamefont {D.}~\bibnamefont
  {Bonamy}}, \bibinfo {author} {\bibfnamefont {S.}~\bibnamefont {Santucci}}, \
  and\ \bibinfo {author} {\bibfnamefont {L.}~\bibnamefont {Ponson}},\
  }\bibfield  {title} {\enquote {\bibinfo {title} {Crackling dynamics in
  material failure as the signature of a self-organized dynamic phase
  transition},}\ }\href {\doibase 10.1103/PhysRevLett.101.045501} {\bibfield
  {journal} {\bibinfo  {journal} {Phys. Rev. Lett.}\ }\textbf {\bibinfo
  {volume} {101}},\ \bibinfo {pages} {045501} (\bibinfo {year}
  {2008})}\BibitemShut {NoStop}%
\bibitem [{\citenamefont {Joanny}\ and\ \citenamefont
  {de~Gennes}(1984)}]{joannydegennes}%
  \BibitemOpen
  \bibfield  {author} {\bibinfo {author} {\bibfnamefont {J.~F.}\ \bibnamefont
  {Joanny}}\ and\ \bibinfo {author} {\bibfnamefont {P.~G.}\ \bibnamefont
  {de~Gennes}},\ }\bibfield  {title} {\enquote {\bibinfo {title} {A model for
  contact angle hysteresis},}\ }\href {\doibase 10.1063/1.447337} {\bibfield
  {journal} {\bibinfo  {journal} {The Journal of Chemical Physics}\ }\textbf
  {\bibinfo {volume} {81}},\ \bibinfo {pages} {552--562} (\bibinfo {year}
  {1984})}\BibitemShut {NoStop}%
\bibitem [{\citenamefont {Moulinet}\ \emph {et~al.}(2004)\citenamefont
  {Moulinet}, \citenamefont {Rosso}, \citenamefont {Krauth},\ and\
  \citenamefont {Rolley}}]{moulinet04}%
  \BibitemOpen
  \bibfield  {author} {\bibinfo {author} {\bibfnamefont {S\'ebastien}\
  \bibnamefont {Moulinet}}, \bibinfo {author} {\bibfnamefont {Alberto}\
  \bibnamefont {Rosso}}, \bibinfo {author} {\bibfnamefont {Werner}\
  \bibnamefont {Krauth}}, \ and\ \bibinfo {author} {\bibfnamefont {Etienne}\
  \bibnamefont {Rolley}},\ }\bibfield  {title} {\enquote {\bibinfo {title}
  {Width distribution of contact lines on a disordered substrate},}\ }\href
  {\doibase 10.1103/PhysRevE.69.035103} {\bibfield  {journal} {\bibinfo
  {journal} {Phys. Rev. E}\ }\textbf {\bibinfo {volume} {69}},\ \bibinfo
  {pages} {035103} (\bibinfo {year} {2004})}\BibitemShut {NoStop}%
\bibitem [{\citenamefont {Doussal}\ \emph {et~al.}(2009)\citenamefont
  {Doussal}, \citenamefont {Wiese}, \citenamefont {Moulinet},\ and\
  \citenamefont {Rolley}}]{LeDoussal2009}%
  \BibitemOpen
  \bibfield  {author} {\bibinfo {author} {\bibfnamefont {P.~Le}\ \bibnamefont
  {Doussal}}, \bibinfo {author} {\bibfnamefont {K.~J.}\ \bibnamefont {Wiese}},
  \bibinfo {author} {\bibfnamefont {S.}~\bibnamefont {Moulinet}}, \ and\
  \bibinfo {author} {\bibfnamefont {E.}~\bibnamefont {Rolley}},\ }\bibfield
  {title} {\enquote {\bibinfo {title} {Height fluctuations of a contact line: A
  direct measurement of the renormalized disorder correlator},}\ }\href
  {\doibase 10.1209/0295-5075/87/56001} {\bibfield  {journal} {\bibinfo
  {journal} {{EPL} (Europhysics Letters)}\ }\textbf {\bibinfo {volume} {87}},\
  \bibinfo {pages} {56001} (\bibinfo {year} {2009})}\BibitemShut {NoStop}%
\bibitem [{\citenamefont {Baret}\ \emph {et~al.}(2002)\citenamefont {Baret},
  \citenamefont {Vandembroucq},\ and\ \citenamefont {Roux}}]{roux}%
  \BibitemOpen
  \bibfield  {author} {\bibinfo {author} {\bibfnamefont {Jean-Christophe}\
  \bibnamefont {Baret}}, \bibinfo {author} {\bibfnamefont {Damien}\
  \bibnamefont {Vandembroucq}}, \ and\ \bibinfo {author} {\bibfnamefont
  {St\'ephane}\ \bibnamefont {Roux}},\ }\bibfield  {title} {\enquote {\bibinfo
  {title} {Extremal model for amorphous media plasticity},}\ }\href {\doibase
  10.1103/PhysRevLett.89.195506} {\bibfield  {journal} {\bibinfo  {journal}
  {Phys. Rev. Lett.}\ }\textbf {\bibinfo {volume} {89}},\ \bibinfo {pages}
  {195506} (\bibinfo {year} {2002})}\BibitemShut {NoStop}%
\bibitem [{\citenamefont {Lin}\ \emph {et~al.}(2014)\citenamefont {Lin},
  \citenamefont {Lerner}, \citenamefont {Rosso},\ and\ \citenamefont
  {Wyart}}]{linjie}%
  \BibitemOpen
  \bibfield  {author} {\bibinfo {author} {\bibfnamefont {Jie}\ \bibnamefont
  {Lin}}, \bibinfo {author} {\bibfnamefont {Edan}\ \bibnamefont {Lerner}},
  \bibinfo {author} {\bibfnamefont {Alberto}\ \bibnamefont {Rosso}}, \ and\
  \bibinfo {author} {\bibfnamefont {Matthieu}\ \bibnamefont {Wyart}},\
  }\bibfield  {title} {\enquote {\bibinfo {title} {Scaling description of the
  yielding transition in soft amorphous solids at zero temperature},}\ }\href
  {\doibase 10.1073/pnas.1406391111} {\bibfield  {journal} {\bibinfo  {journal}
  {Proceedings of the National Academy of Sciences}\ }\textbf {\bibinfo
  {volume} {111}},\ \bibinfo {pages} {14382--14387} (\bibinfo {year}
  {2014})}\BibitemShut {NoStop}%
\bibitem [{\citenamefont {Hallatschek}\ and\ \citenamefont
  {Fisher}(2014)}]{doi:10.1073/pnas.1404663111}%
  \BibitemOpen
  \bibfield  {author} {\bibinfo {author} {\bibfnamefont {Oskar}\ \bibnamefont
  {Hallatschek}}\ and\ \bibinfo {author} {\bibfnamefont {Daniel~S.}\
  \bibnamefont {Fisher}},\ }\bibfield  {title} {\enquote {\bibinfo {title}
  {Acceleration of evolutionary spread by long-range dispersal},}\ }\href
  {\doibase 10.1073/pnas.1404663111} {\bibfield  {journal} {\bibinfo  {journal}
  {Proceedings of the National Academy of Sciences}\ }\textbf {\bibinfo
  {volume} {111}},\ \bibinfo {pages} {E4911--E4919} (\bibinfo {year}
  {2014})}\BibitemShut {NoStop}%
\bibitem [{\citenamefont {Chatterjee}\ and\ \citenamefont
  {S.~Dey}(2016)}]{chtterjee}%
  \BibitemOpen
  \bibfield  {author} {\bibinfo {author} {\bibfnamefont {Shirshendu}\
  \bibnamefont {Chatterjee}}\ and\ \bibinfo {author} {\bibfnamefont {Partha}\
  \bibnamefont {S.~Dey}},\ }\bibfield  {title} {\enquote {\bibinfo {title}
  {Multiple phase transitions in long-range first-passage percolation on square
  lattices},}\ }\href {\doibase https://doi.org/10.1002/cpa.21571} {\bibfield
  {journal} {\bibinfo  {journal} {Communications on Pure and Applied
  Mathematics}\ }\textbf {\bibinfo {volume} {69}},\ \bibinfo {pages} {203--256}
  (\bibinfo {year} {2016})}\BibitemShut {NoStop}%
\bibitem [{\citenamefont {Cao}\ \emph {et~al.}(2017)\citenamefont {Cao},
  \citenamefont {Rosso}, \citenamefont {Bouchaud},\ and\ \citenamefont
  {Le~Doussal}}]{caobouchaud}%
  \BibitemOpen
  \bibfield  {author} {\bibinfo {author} {\bibfnamefont {Xiangyu}\ \bibnamefont
  {Cao}}, \bibinfo {author} {\bibfnamefont {Alberto}\ \bibnamefont {Rosso}},
  \bibinfo {author} {\bibfnamefont {Jean-Philippe}\ \bibnamefont {Bouchaud}}, \
  and\ \bibinfo {author} {\bibfnamefont {Pierre}\ \bibnamefont {Le~Doussal}},\
  }\bibfield  {title} {\enquote {\bibinfo {title} {Genuine localization
  transition in a long-range hopping model},}\ }\href {\doibase
  10.1103/PhysRevE.95.062118} {\bibfield  {journal} {\bibinfo  {journal} {Phys.
  Rev. E}\ }\textbf {\bibinfo {volume} {95}},\ \bibinfo {pages} {062118}
  (\bibinfo {year} {2017})}\BibitemShut {NoStop}%
\bibitem [{\citenamefont {Hinrichsen}(2000)}]{hinrichsen}%
  \BibitemOpen
  \bibfield  {author} {\bibinfo {author} {\bibfnamefont {Haye}\ \bibnamefont
  {Hinrichsen}},\ }\bibfield  {title} {\enquote {\bibinfo {title}
  {Non-equilibrium critical phenomena and phase transitions into absorbing
  states},}\ }\href {\doibase 10.1080/00018730050198152} {\bibfield  {journal}
  {\bibinfo  {journal} {Advances in Physics}\ }\textbf {\bibinfo {volume}
  {49}},\ \bibinfo {pages} {815--958} (\bibinfo {year} {2000})}\BibitemShut
  {NoStop}%
\bibitem [{\citenamefont {Janssen}\ and\ \citenamefont
  {Stenull}(2008)}]{jansen08DP}%
  \BibitemOpen
  \bibfield  {author} {\bibinfo {author} {\bibfnamefont {Hans-Karl}\
  \bibnamefont {Janssen}}\ and\ \bibinfo {author} {\bibfnamefont {Olaf}\
  \bibnamefont {Stenull}},\ }\bibfield  {title} {\enquote {\bibinfo {title}
  {Field theory of directed percolation with long-range spreading},}\ }\href
  {\doibase 10.1103/PhysRevE.78.061117} {\bibfield  {journal} {\bibinfo
  {journal} {Phys. Rev. E}\ }\textbf {\bibinfo {volume} {78}},\ \bibinfo
  {pages} {061117} (\bibinfo {year} {2008})}\BibitemShut {NoStop}%
\bibitem [{\citenamefont {Grassberger}(2013{\natexlab{a}})}]{Grassberger13-1d}%
  \BibitemOpen
  \bibfield  {author} {\bibinfo {author} {\bibfnamefont {Peter}\ \bibnamefont
  {Grassberger}},\ }\bibfield  {title} {\enquote {\bibinfo {title} {{SIR}
  epidemics with long-range infection in one dimension},}\ }\href {\doibase
  10.1088/1742-5468/2013/04/p04004} {\bibfield  {journal} {\bibinfo  {journal}
  {Journal of Statistical Mechanics: Theory and Experiment}\ }\textbf {\bibinfo
  {volume} {2013}},\ \bibinfo {pages} {P04004} (\bibinfo {year}
  {2013}{\natexlab{a}})}\BibitemShut {NoStop}%
\bibitem [{\citenamefont {Grassberger}(2013{\natexlab{b}})}]{grassberger13-2d}%
  \BibitemOpen
  \bibfield  {author} {\bibinfo {author} {\bibfnamefont {Peter}\ \bibnamefont
  {Grassberger}},\ }\bibfield  {title} {\enquote {\bibinfo {title}
  {Two-dimensional sir epidemics with long range infection},}\ }\href@noop {}
  {\bibfield  {journal} {\bibinfo  {journal} {Journal of Statistical Physics}\
  }\textbf {\bibinfo {volume} {153}},\ \bibinfo {pages} {289--311} (\bibinfo
  {year} {2013}{\natexlab{b}})}\BibitemShut {NoStop}%
\bibitem [{SM()}]{SM}%
  \BibitemOpen
  \href@noop {} {}\bibinfo {note} {See Supplemental Material.}\BibitemShut
  {Stop}%
\bibitem [{pie(2022{\natexlab{a}})}]{pierrenew}%
  \BibitemOpen
  \href@noop {} {\bibfield  {journal} {\bibinfo  {journal} {For a recent review
  see Pierre Le Doussal, {\it Equivalence of mean-field avalanches and
  branching diffusions: From the Brownian force model to the super-Brownian
  motion}, ArXiv:2203.10512}\ } (\bibinfo {year}
  {2022}{\natexlab{a}})}\BibitemShut {NoStop}%
\bibitem [{\citenamefont {Doussal}\ and\ \citenamefont {Wiese}(2012)}]{pldw12}%
  \BibitemOpen
  \bibfield  {author} {\bibinfo {author} {\bibfnamefont {P.~Le}\ \bibnamefont
  {Doussal}}\ and\ \bibinfo {author} {\bibfnamefont {K.~J.}\ \bibnamefont
  {Wiese}},\ }\bibfield  {title} {\enquote {\bibinfo {title} {Distribution of
  velocities in an avalanche},}\ }\href {\doibase 10.1209/0295-5075/97/46004}
  {\bibfield  {journal} {\bibinfo  {journal} {{EPL} (Europhysics Letters)}\
  }\textbf {\bibinfo {volume} {97}},\ \bibinfo {pages} {46004} (\bibinfo {year}
  {2012})}\BibitemShut {NoStop}%
\bibitem [{\citenamefont {Le~Doussal}\ and\ \citenamefont
  {Wiese}(2013)}]{pldw13}%
  \BibitemOpen
  \bibfield  {author} {\bibinfo {author} {\bibfnamefont {Pierre}\ \bibnamefont
  {Le~Doussal}}\ and\ \bibinfo {author} {\bibfnamefont {Kay~J\"org}\
  \bibnamefont {Wiese}},\ }\bibfield  {title} {\enquote {\bibinfo {title}
  {Avalanche dynamics of elastic interfaces},}\ }\href {\doibase
  10.1103/PhysRevE.88.022106} {\bibfield  {journal} {\bibinfo  {journal} {Phys.
  Rev. E}\ }\textbf {\bibinfo {volume} {88}},\ \bibinfo {pages} {022106}
  (\bibinfo {year} {2013})}\BibitemShut {NoStop}%
\bibitem [{\citenamefont {M\aa{}l\o{}y}\ \emph {et~al.}(2006)\citenamefont
  {M\aa{}l\o{}y}, \citenamefont {Santucci}, \citenamefont {Schmittbuhl},\ and\
  \citenamefont {Toussaint}}]{toussaint06}%
  \BibitemOpen
  \bibfield  {author} {\bibinfo {author} {\bibfnamefont {Knut~J\o{}rgen}\
  \bibnamefont {M\aa{}l\o{}y}}, \bibinfo {author} {\bibfnamefont {St\'ephane}\
  \bibnamefont {Santucci}}, \bibinfo {author} {\bibfnamefont {Jean}\
  \bibnamefont {Schmittbuhl}}, \ and\ \bibinfo {author} {\bibfnamefont
  {Renaud}\ \bibnamefont {Toussaint}},\ }\bibfield  {title} {\enquote {\bibinfo
  {title} {Local waiting time fluctuations along a randomly pinned crack
  front},}\ }\href {\doibase 10.1103/PhysRevLett.96.045501} {\bibfield
  {journal} {\bibinfo  {journal} {Phys. Rev. Lett.}\ }\textbf {\bibinfo
  {volume} {96}},\ \bibinfo {pages} {045501} (\bibinfo {year}
  {2006})}\BibitemShut {NoStop}%
\bibitem [{\citenamefont {Laurson}\ \emph {et~al.}(2010)\citenamefont
  {Laurson}, \citenamefont {Santucci},\ and\ \citenamefont
  {Zapperi}}]{zapperi10}%
  \BibitemOpen
  \bibfield  {author} {\bibinfo {author} {\bibfnamefont {Lasse}\ \bibnamefont
  {Laurson}}, \bibinfo {author} {\bibfnamefont {Stephane}\ \bibnamefont
  {Santucci}}, \ and\ \bibinfo {author} {\bibfnamefont {Stefano}\ \bibnamefont
  {Zapperi}},\ }\bibfield  {title} {\enquote {\bibinfo {title} {Avalanches and
  clusters in planar crack front propagation},}\ }\href {\doibase
  10.1103/PhysRevE.81.046116} {\bibfield  {journal} {\bibinfo  {journal} {Phys.
  Rev. E}\ }\textbf {\bibinfo {volume} {81}},\ \bibinfo {pages} {046116}
  (\bibinfo {year} {2010})}\BibitemShut {NoStop}%
\bibitem [{\citenamefont {Le~Priol}\ \emph {et~al.}(2021)\citenamefont
  {Le~Priol}, \citenamefont {Le~Doussal},\ and\ \citenamefont
  {Rosso}}]{lepriol21}%
  \BibitemOpen
  \bibfield  {author} {\bibinfo {author} {\bibfnamefont {Cl\'ement}\
  \bibnamefont {Le~Priol}}, \bibinfo {author} {\bibfnamefont {Pierre}\
  \bibnamefont {Le~Doussal}}, \ and\ \bibinfo {author} {\bibfnamefont
  {Alberto}\ \bibnamefont {Rosso}},\ }\bibfield  {title} {\enquote {\bibinfo
  {title} {Spatial clustering of depinning avalanches in presence of long-range
  interactions},}\ }\href {\doibase 10.1103/PhysRevLett.126.025702} {\bibfield
  {journal} {\bibinfo  {journal} {Phys. Rev. Lett.}\ }\textbf {\bibinfo
  {volume} {126}},\ \bibinfo {pages} {025702} (\bibinfo {year}
  {2021})}\BibitemShut {NoStop}%
\bibitem [{\citenamefont {Erta\ifmmode~\mbox{\c{s}}\else \c{s}\fi{}}\ and\
  \citenamefont {Kardar}(1994)}]{kardar94}%
  \BibitemOpen
  \bibfield  {author} {\bibinfo {author} {\bibfnamefont {Deniz}\ \bibnamefont
  {Erta\ifmmode~\mbox{\c{s}}\else \c{s}\fi{}}}\ and\ \bibinfo {author}
  {\bibfnamefont {Mehran}\ \bibnamefont {Kardar}},\ }\bibfield  {title}
  {\enquote {\bibinfo {title} {Critical dynamics of contact line depinning},}\
  }\href {\doibase 10.1103/PhysRevE.49.R2532} {\bibfield  {journal} {\bibinfo
  {journal} {Phys. Rev. E}\ }\textbf {\bibinfo {volume} {49}},\ \bibinfo
  {pages} {R2532--R2535} (\bibinfo {year} {1994})}\BibitemShut {NoStop}%
\bibitem [{\citenamefont {Le~Doussal}\ \emph {et~al.}(2002)\citenamefont
  {Le~Doussal}, \citenamefont {Wiese},\ and\ \citenamefont {Chauve}}]{twoloop}%
  \BibitemOpen
  \bibfield  {author} {\bibinfo {author} {\bibfnamefont {Pierre}\ \bibnamefont
  {Le~Doussal}}, \bibinfo {author} {\bibfnamefont {Kay~J\"org}\ \bibnamefont
  {Wiese}}, \ and\ \bibinfo {author} {\bibfnamefont {Pascal}\ \bibnamefont
  {Chauve}},\ }\bibfield  {title} {\enquote {\bibinfo {title} {Two-loop
  functional renormalization group theory of the depinning transition},}\
  }\href {\doibase 10.1103/PhysRevB.66.174201} {\bibfield  {journal} {\bibinfo
  {journal} {Phys. Rev. B}\ }\textbf {\bibinfo {volume} {66}},\ \bibinfo
  {pages} {174201} (\bibinfo {year} {2002})}\BibitemShut {NoStop}%
\bibitem [{\citenamefont {Rosso}\ and\ \citenamefont
  {Krauth}(2002)}]{rosso02rough}%
  \BibitemOpen
  \bibfield  {author} {\bibinfo {author} {\bibfnamefont {Alberto}\ \bibnamefont
  {Rosso}}\ and\ \bibinfo {author} {\bibfnamefont {Werner}\ \bibnamefont
  {Krauth}},\ }\bibfield  {title} {\enquote {\bibinfo {title} {Roughness at the
  depinning threshold for a long-range elastic string},}\ }\href {\doibase
  10.1103/PhysRevE.65.025101} {\bibfield  {journal} {\bibinfo  {journal} {Phys.
  Rev. E}\ }\textbf {\bibinfo {volume} {65}},\ \bibinfo {pages} {025101}
  (\bibinfo {year} {2002})}\BibitemShut {NoStop}%
\bibitem [{\citenamefont {Thiery}\ \emph {et~al.}(2015)\citenamefont {Thiery},
  \citenamefont {Doussal},\ and\ \citenamefont {Wiese}}]{Thiery_2015}%
  \BibitemOpen
  \bibfield  {author} {\bibinfo {author} {\bibfnamefont {Thimoth{\'{e}}e}\
  \bibnamefont {Thiery}}, \bibinfo {author} {\bibfnamefont {Pierre~Le}\
  \bibnamefont {Doussal}}, \ and\ \bibinfo {author} {\bibfnamefont {Kay~Jörg}\
  \bibnamefont {Wiese}},\ }\bibfield  {title} {\enquote {\bibinfo {title}
  {Spatial shape of avalanches in the brownian force model},}\ }\href {\doibase
  10.1088/1742-5468/2015/08/p08019} {\bibfield  {journal} {\bibinfo  {journal}
  {Journal of Statistical Mechanics: Theory and Experiment}\ }\textbf {\bibinfo
  {volume} {2015}},\ \bibinfo {pages} {P08019} (\bibinfo {year}
  {2015})}\BibitemShut {NoStop}%
\bibitem [{Note1()}]{Note1}%
  \BibitemOpen
  \bibinfo {note} {A similar but less precise description applies to 2D if we
  replace a gap of length $g$ by an ``empty space'' of area $g^2$.}\BibitemShut
  {Stop}%
\bibitem [{\citenamefont {Fisher}(1937)}]{fisher}%
  \BibitemOpen
  \bibfield  {author} {\bibinfo {author} {\bibfnamefont {R.~A.}\ \bibnamefont
  {Fisher}},\ }\bibfield  {title} {\enquote {\bibinfo {title} {The wave of
  advance of advantageous genes},}\ }\href {\doibase
  https://doi.org/10.1111/j.1469-1809.1937.tb02153.x} {\bibfield  {journal}
  {\bibinfo  {journal} {Annals of Eugenics}\ }\textbf {\bibinfo {volume} {7}},\
  \bibinfo {pages} {355--369} (\bibinfo {year} {1937})}\BibitemShut {NoStop}%
\bibitem [{\citenamefont {Kolmogorov}\ \emph {et~al.}(1937)\citenamefont
  {Kolmogorov}, \citenamefont {Petrovsky},\ and\ \citenamefont
  {Piscounov}}]{kpp}%
  \BibitemOpen
  \bibfield  {author} {\bibinfo {author} {\bibfnamefont {A.}~\bibnamefont
  {Kolmogorov}}, \bibinfo {author} {\bibfnamefont {I.}~\bibnamefont
  {Petrovsky}}, \ and\ \bibinfo {author} {\bibfnamefont {N.}~\bibnamefont
  {Piscounov}},\ }\bibfield  {title} {\enquote {\bibinfo {title} {Etude de
  l'\'equation de la diffusion avec croissance de la quantit \'e de mati\`ere
  et son application \`a un probl\`eme biologique},}\ }\href@noop {} {\bibfield
   {journal} {\bibinfo  {journal} {Bull. Univ. Etat Moscou A}\ }\textbf
  {\bibinfo {volume} {1}},\ \bibinfo {pages} {1} (\bibinfo {year}
  {1937})}\BibitemShut {NoStop}%
\bibitem [{\citenamefont {Dawson}(1975)}]{dawson}%
  \BibitemOpen
  \bibfield  {author} {\bibinfo {author} {\bibfnamefont {D.~A.}\ \bibnamefont
  {Dawson}},\ }\bibfield  {title} {\enquote {\bibinfo {title} {Stochastic
  evolution equations and related measure processes},}\ }\href {\doibase
  https://doi.org/10.1016/0047-259X(75)90054-8} {\bibfield  {journal} {\bibinfo
   {journal} {Journal of Multivariate Analysis}\ }\textbf {\bibinfo {volume}
  {5}},\ \bibinfo {pages} {1--52} (\bibinfo {year} {1975})}\BibitemShut
  {NoStop}%
\bibitem [{\citenamefont {Watanabe}(1968)}]{wanatabe}%
  \BibitemOpen
  \bibfield  {author} {\bibinfo {author} {\bibfnamefont {Shinzo}\ \bibnamefont
  {Watanabe}},\ }\bibfield  {title} {\enquote {\bibinfo {title} {{A limit
  theorem of branching processes and continuous state branching processes}},}\
  }\href {\doibase 10.1215/kjm/1250524180} {\bibfield  {journal} {\bibinfo
  {journal} {Journal of Mathematics of Kyoto University}\ }\textbf {\bibinfo
  {volume} {8}},\ \bibinfo {pages} {141 -- 167} (\bibinfo {year}
  {1968})}\BibitemShut {NoStop}%
\bibitem [{\citenamefont {Cabr{\'e}}\ and\ \citenamefont
  {Roquejoffre}(2013)}]{Cabre13KPP}%
  \BibitemOpen
  \bibfield  {author} {\bibinfo {author} {\bibfnamefont {Xavier}\ \bibnamefont
  {Cabr{\'e}}}\ and\ \bibinfo {author} {\bibfnamefont {Jean-Michel}\
  \bibnamefont {Roquejoffre}},\ }\bibfield  {title} {\enquote {\bibinfo {title}
  {{The Influence of Fractional Diffusion in Fisher-KPP Equations}},}\ }\href
  {\doibase 10.1007/s00220-013-1682-5} {\bibfield  {journal} {\bibinfo
  {journal} {Communications in Mathematical Physics}\ }\textbf {\bibinfo
  {volume} {320}},\ \bibinfo {pages} {679--722} (\bibinfo {year}
  {2013})}\BibitemShut {NoStop}%
\bibitem [{\citenamefont {Le~Priol}(2020)}]{lepriol-thesis}%
  \BibitemOpen
  \bibfield  {author} {\bibinfo {author} {\bibfnamefont {Cl{\'e}ment}\
  \bibnamefont {Le~Priol}},\ }\emph {\bibinfo {title} {{Long-range interactions
  in the avalanches of elastic interfaces}}},\ \href
  {https://tel.archives-ouvertes.fr/tel-03191605} {\bibinfo {type} {Theses}},\
  \bibinfo  {school} {{Universit{\'e} Paris sciences et lettres}} (\bibinfo
  {year} {2020})\BibitemShut {NoStop}%
\bibitem [{\citenamefont {Mistry}\ \emph {et~al.}(2021)\citenamefont {Mistry},
  \citenamefont {Litvinova}, \citenamefont {Pastore~y Piontti}, \citenamefont
  {Chinazzi}, \citenamefont {Fumanelli}, \citenamefont {Gomes}, \citenamefont
  {Haque}, \citenamefont {Liu}, \citenamefont {Mu}, \citenamefont {Xiong},
  \citenamefont {Halloran}, \citenamefont {Longini}, \citenamefont {Merler},
  \citenamefont {Ajelli},\ and\ \citenamefont {Vespignani}}]{mistry21}%
  \BibitemOpen
  \bibfield  {author} {\bibinfo {author} {\bibfnamefont {Dina}\ \bibnamefont
  {Mistry}}, \bibinfo {author} {\bibfnamefont {Maria}\ \bibnamefont
  {Litvinova}}, \bibinfo {author} {\bibfnamefont {Ana}\ \bibnamefont {Pastore~y
  Piontti}}, \bibinfo {author} {\bibfnamefont {Matteo}\ \bibnamefont
  {Chinazzi}}, \bibinfo {author} {\bibfnamefont {Laura}\ \bibnamefont
  {Fumanelli}}, \bibinfo {author} {\bibfnamefont {Marcelo F.~C.}\ \bibnamefont
  {Gomes}}, \bibinfo {author} {\bibfnamefont {Syed~A.}\ \bibnamefont {Haque}},
  \bibinfo {author} {\bibfnamefont {Quan-Hui}\ \bibnamefont {Liu}}, \bibinfo
  {author} {\bibfnamefont {Kunpeng}\ \bibnamefont {Mu}}, \bibinfo {author}
  {\bibfnamefont {Xinyue}\ \bibnamefont {Xiong}}, \bibinfo {author}
  {\bibfnamefont {M.~Elizabeth}\ \bibnamefont {Halloran}}, \bibinfo {author}
  {\bibfnamefont {Ira~M.}\ \bibnamefont {Longini}}, \bibinfo {author}
  {\bibfnamefont {Stefano}\ \bibnamefont {Merler}}, \bibinfo {author}
  {\bibfnamefont {Marco}\ \bibnamefont {Ajelli}}, \ and\ \bibinfo {author}
  {\bibfnamefont {Alessandro}\ \bibnamefont {Vespignani}},\ }\bibfield  {title}
  {\enquote {\bibinfo {title} {Inferring high-resolution human mixing patterns
  for disease modeling},}\ }\href {\doibase 10.1038/s41467-020-20544-y}
  {\bibfield  {journal} {\bibinfo  {journal} {Nature Communications}\ }\textbf
  {\bibinfo {volume} {12}},\ \bibinfo {pages} {323} (\bibinfo {year}
  {2021})}\BibitemShut {NoStop}%
\bibitem [{\citenamefont {Lloyd-Smith}\ \emph {et~al.}(2005)\citenamefont
  {Lloyd-Smith}, \citenamefont {Schreiber}, \citenamefont {Kopp},\ and\
  \citenamefont {Getz}}]{lloyd05}%
  \BibitemOpen
  \bibfield  {author} {\bibinfo {author} {\bibfnamefont {J.~O.}\ \bibnamefont
  {Lloyd-Smith}}, \bibinfo {author} {\bibfnamefont {S.~J.}\ \bibnamefont
  {Schreiber}}, \bibinfo {author} {\bibfnamefont {P.~E.}\ \bibnamefont {Kopp}},
  \ and\ \bibinfo {author} {\bibfnamefont {W.~M.}\ \bibnamefont {Getz}},\
  }\bibfield  {title} {\enquote {\bibinfo {title} {Superspreading and the
  effect of individual variation on disease emergence},}\ }\href {\doibase
  10.1038/nature04153} {\bibfield  {journal} {\bibinfo  {journal} {Nature}\
  }\textbf {\bibinfo {volume} {438}},\ \bibinfo {pages} {355--359} (\bibinfo
  {year} {2005})}\BibitemShut {NoStop}%
\bibitem [{\citenamefont {Meyer}\ \emph {et~al.}(1996)\citenamefont {Meyer},
  \citenamefont {Havlin},\ and\ \citenamefont {Bunde}}]{meyer96}%
  \BibitemOpen
  \bibfield  {author} {\bibinfo {author} {\bibfnamefont {Martin}\ \bibnamefont
  {Meyer}}, \bibinfo {author} {\bibfnamefont {Shlomo}\ \bibnamefont {Havlin}},
  \ and\ \bibinfo {author} {\bibfnamefont {Armin}\ \bibnamefont {Bunde}},\
  }\bibfield  {title} {\enquote {\bibinfo {title} {Clustering of independently
  diffusing individuals by birth and death processes},}\ }\href {\doibase
  10.1103/PhysRevE.54.5567} {\bibfield  {journal} {\bibinfo  {journal} {Phys.
  Rev. E}\ }\textbf {\bibinfo {volume} {54}},\ \bibinfo {pages} {5567--5570}
  (\bibinfo {year} {1996})}\BibitemShut {NoStop}%
\bibitem [{\citenamefont {Houchmandzadeh}(2008)}]{Houchmandzadeh}%
  \BibitemOpen
  \bibfield  {author} {\bibinfo {author} {\bibfnamefont {B.}~\bibnamefont
  {Houchmandzadeh}},\ }\bibfield  {title} {\enquote {\bibinfo {title} {Neutral
  clustering in a simple experimental ecological community},}\ }\href {\doibase
  10.1103/PhysRevLett.101.078103} {\bibfield  {journal} {\bibinfo  {journal}
  {Phys. Rev. Lett.}\ }\textbf {\bibinfo {volume} {101}},\ \bibinfo {pages}
  {078103} (\bibinfo {year} {2008})}\BibitemShut {NoStop}%
\bibitem [{\citenamefont {Marro}\ and\ \citenamefont
  {Dickman}(1999)}]{marro_dickman_1999}%
  \BibitemOpen
  \bibfield  {author} {\bibinfo {author} {\bibfnamefont {Joaquin}\ \bibnamefont
  {Marro}}\ and\ \bibinfo {author} {\bibfnamefont {Ronald}\ \bibnamefont
  {Dickman}},\ }\href {\doibase 10.1017/CBO9780511524288} {\emph {\bibinfo
  {title} {Nonequilibrium Phase Transitions in Lattice Models}}},\ Collection
  Alea-Saclay: Monographs and Texts in Statistical Physics\ (\bibinfo
  {publisher} {Cambridge University Press},\ \bibinfo {year}
  {1999})\BibitemShut {NoStop}%
\bibitem [{\citenamefont {Krapivsky}\ \emph {et~al.}(2010)\citenamefont
  {Krapivsky}, \citenamefont {Redner},\ and\ \citenamefont
  {Ben-Naim}}]{krapivsky2010kinetic}%
  \BibitemOpen
  \bibfield  {author} {\bibinfo {author} {\bibfnamefont {Pavel~L}\ \bibnamefont
  {Krapivsky}}, \bibinfo {author} {\bibfnamefont {Sidney}\ \bibnamefont
  {Redner}}, \ and\ \bibinfo {author} {\bibfnamefont {Eli}\ \bibnamefont
  {Ben-Naim}},\ }\href@noop {} {\emph {\bibinfo {title} {A kinetic view of
  statistical physics}}}\ (\bibinfo  {publisher} {Cambridge University Press},\
  \bibinfo {year} {2010})\BibitemShut {NoStop}%
\bibitem [{pie(2022{\natexlab{b}})}]{pierrenewBFM}%
  \BibitemOpen
  \href@noop {} {\bibfield  {journal} {\bibinfo  {journal} {Pierre Le Doussal,
  {\it More on the Brownian force model: avalanche shapes, tip driven, higher
  d}, ArXiv:2203.10544}\ } (\bibinfo {year} {2022}{\natexlab{b}})}\BibitemShut
  {NoStop}%
\bibitem [{\citenamefont {Mytnik}\ and\ \citenamefont
  {Perkins}(2019)}]{MytnikBoundary}%
  \BibitemOpen
  \bibfield  {author} {\bibinfo {author} {\bibfnamefont {Leonid}\ \bibnamefont
  {Mytnik}}\ and\ \bibinfo {author} {\bibfnamefont {Edwin}\ \bibnamefont
  {Perkins}},\ }\bibfield  {title} {\enquote {\bibinfo {title} {The dimension
  of the boundary of super-brownian motion},}\ }\href {\doibase
  10.1007/s00440-018-0866-5} {\bibfield  {journal} {\bibinfo  {journal}
  {Probability Theory and Related Fields}\ }\textbf {\bibinfo {volume} {174}},\
  \bibinfo {pages} {821--885} (\bibinfo {year} {2019})}\BibitemShut {NoStop}%
\bibitem [{\citenamefont {Narayan}\ and\ \citenamefont
  {Fisher}(1993)}]{narayan}%
  \BibitemOpen
  \bibfield  {author} {\bibinfo {author} {\bibfnamefont {Onuttom}\ \bibnamefont
  {Narayan}}\ and\ \bibinfo {author} {\bibfnamefont {Daniel~S.}\ \bibnamefont
  {Fisher}},\ }\bibfield  {title} {\enquote {\bibinfo {title} {Threshold
  critical dynamics of driven interfaces in random media},}\ }\href {\doibase
  10.1103/PhysRevB.48.7030} {\bibfield  {journal} {\bibinfo  {journal} {Phys.
  Rev. B}\ }\textbf {\bibinfo {volume} {48}},\ \bibinfo {pages} {7030--7042}
  (\bibinfo {year} {1993})}\BibitemShut {NoStop}%
\bibitem [{\citenamefont {Zapperi}\ \emph {et~al.}(1998)\citenamefont
  {Zapperi}, \citenamefont {Cizeau}, \citenamefont {Durin},\ and\ \citenamefont
  {Stanley}}]{zapperi98}%
  \BibitemOpen
  \bibfield  {author} {\bibinfo {author} {\bibfnamefont {Stefano}\ \bibnamefont
  {Zapperi}}, \bibinfo {author} {\bibfnamefont {Pierre}\ \bibnamefont
  {Cizeau}}, \bibinfo {author} {\bibfnamefont {Gianfranco}\ \bibnamefont
  {Durin}}, \ and\ \bibinfo {author} {\bibfnamefont {H.~Eugene}\ \bibnamefont
  {Stanley}},\ }\bibfield  {title} {\enquote {\bibinfo {title} {Dynamics of a
  ferromagnetic domain wall: Avalanches, depinning transition, and the
  barkhausen effect},}\ }\href {\doibase 10.1103/PhysRevB.58.6353} {\bibfield
  {journal} {\bibinfo  {journal} {Phys. Rev. B}\ }\textbf {\bibinfo {volume}
  {58}},\ \bibinfo {pages} {6353--6366} (\bibinfo {year} {1998})}\BibitemShut
  {NoStop}%
\bibitem [{\citenamefont {Dobrinevski}\ \emph {et~al.}(2014)\citenamefont
  {Dobrinevski}, \citenamefont {Doussal},\ and\ \citenamefont
  {Wiese}}]{DobrinevskiLeDoussalWiese2014a}%
  \BibitemOpen
  \bibfield  {author} {\bibinfo {author} {\bibfnamefont {Alexander}\
  \bibnamefont {Dobrinevski}}, \bibinfo {author} {\bibfnamefont {Pierre~Le}\
  \bibnamefont {Doussal}}, \ and\ \bibinfo {author} {\bibfnamefont {Kay~Jörg}\
  \bibnamefont {Wiese}},\ }\bibfield  {title} {\enquote {\bibinfo {title}
  {Avalanche shape and exponents beyond mean-field theory},}\ }\href {\doibase
  10.1209/0295-5075/108/66002} {\bibfield  {journal} {\bibinfo  {journal}
  {{EPL} (Europhysics Letters)}\ }\textbf {\bibinfo {volume} {108}},\ \bibinfo
  {pages} {66002} (\bibinfo {year} {2014})}\BibitemShut {NoStop}%
\end{thebibliography}%

\end{document}